\documentclass[12pt]{article}
\usepackage[utf8]{inputenc}
\usepackage{authblk}
\usepackage{amsmath}
\usepackage{graphicx}
\graphicspath{{./images/} }
\usepackage{physics}
\usepackage{xcolor}
\usepackage[english]{babel}
\usepackage{makecell}
\usepackage{float}
\usepackage{geometry}
\usepackage{csquotes}
\usepackage{cite}
\usepackage{hyperref}
\usepackage{caption}
\usepackage{hhline}
\hypersetup{
    colorlinks,
    citecolor=black,
    filecolor=black,
    linkcolor=black,
    urlcolor=black
}

\title{A lab-based test of the gravitational redshift with a miniature clock network}

\author{Xin Zheng$^{1,\dagger}$, Jonathan Dolde$^{1,\dagger}$, Matthew C. Cambria$^{1}$, Hong Ming Lim$^{1}$, Shimon Kolkowitz$^{1,2,\ast}$}

\affil{\textit{\normalsize${}^{1}$Department of Physics, University of Wisconsin-Madison, WI 53706, USA}}
\affil{\textit{\normalsize${}^{2}$Department of Physics, University of California, Berkeley, CA 94720, USA}}

\affil{\textit{\normalsize${}^{\dagger}$These authors contributed equally to this work}}

\affil{\textit{\normalsize${}^{\ast}$To whom correspondence should be addressed; E-mail: kolkowitz@berkeley.edu}}

\date{}

\begin{document}

\baselineskip24pt

\renewcommand{\abstractname}{\vspace{-\baselineskip}} % remove Abstract title

\maketitle

\begin{abstract}
    \baselineskip24pt
    \textbf{
    Einstein's theory of general relativity predicts that 
    a clock at a higher gravitational potential will tick faster than an otherwise identical clock at a lower potential, an effect known as the gravitational redshift.
    Here we perform a laboratory-based, blinded test of the gravitational redshift using differential clock comparisons within an evenly spaced array of 5 atomic ensembles spanning a height difference of 1 cm.
    We measure a fractional frequency gradient of $[-12.4\pm0.7_{\rm{(stat)}}\pm2.5_{\rm{(sys)}}]\times10^{-19}/$cm, consistent with the expected 
    redshift gradient of $-10.9\times10^{-19}/$cm.
    Our results can also be viewed as relativistic gravitational potential difference measurements with sensitivity to mm scale changes in height on the surface of the Earth.
    These results highlight the potential of local-oscillator-independent differential clock comparisons for emerging applications of optical atomic clocks including geodesy, 
    searches for new physics, 
    gravitational wave detection, 
    and explorations of the interplay between quantum mechanics and gravity.
    }
\end{abstract}

\section*{Introduction}

Einstein’s theory of general relativity~\cite{Einstein_GeneralRelativity_1915} has thus far been consistent with every experimental test performed~\cite{Will_GR_2014},
including classical~\cite{DeflectionBySun_Dyson_1920,MercuryPrecession_RMP_1947}, modern~\cite{PoundRebka_1959_PRL,ciufolini_confirmation_2004} and strong field cosmological tests~\cite{stairs_testing_2003,GW_observation_2016,archibald_universality_2018}. 
However, despite the successful integration of special relativity and quantum mechanics as quantum field theory,
there is currently no theory that successfully unifies general relativity with quantum mechanics.
This motivates continued experimental tests at new length scales, and suggests that performing precision tests of general relativity with quantum systems may offer a way to explore the interplay between general relativity and quantum mechanics~\cite{entanglement_gravity_1975,indirect_quantumgravity_1981,Zych_Proper_Time}.

The gravitational redshift is a central prediction of general relativity. Thanks to rapid advancements in their stability and accuracy~\cite{ludlow_optical_2015,SingleIon_-18_2016,mcgrew_atomic_2018,bothwell_jila_2019,Brewer_IonClock_2019,Oelker_ClockStability_2019,campbell_fermi_degenerate_2017},
atomic clocks have now enabled tests of the gravitational redshift over a wide range of length scales~\cite{Hafele_AroundTheWorld_1972,chou_optical_2010,Redshift_Eccentric_2018,Redshift_Eccentric2_2018,takamoto_test_2020,bothwell_resolving_2022}.
Recent tests of the gravitational redshift include a frequency comparison between two single-ion clocks with one of the clocks elevated by 30 cm~\cite{chou_optical_2010}, comparisons between terrestrial clocks and microwave atomic clocks in eccentric orbits which have produced the strongest limits on deviations from the expected redshift~\cite{Redshift_Eccentric_2018,Redshift_Eccentric2_2018},
a frequency comparison between two synchronously linked portable ${}^{87}$Sr optical lattice clocks with a 450 m height difference at the Tokyo Skytree tower~\cite{takamoto_test_2020}
resulting in the most precise terrestrial constraint on deviations from the gravitational redshift at the $10^{-5}$ level, and a recent in-situ synchronous frequency gradient measurement across a millimeter-scale atomic ensemble with an unprecedented differential precision of $7.6\times10^{-21}$~\cite{bothwell_resolving_2022}.

For two otherwise identical clocks experiencing the same gravitational field with a height difference $\Delta h$, 
their frequency difference $\delta f$ due to the gravitational redshift is given by
\begin{equation}
    \frac{\delta f}{f} \approx \frac{g\Delta h}{c^2}
    \label{rs}
\end{equation}
where $f$ is the clock frequency,
$c$ is the speed of light,
and $g$ is the gravitational acceleration.
Near the surface of Earth,
this amounts to a fractional frequency shift of $1.1\times10^{-18}$ per centimeter of vertical displacement.
With optical clocks now reaching instabilities and inaccuracies at the level of $10^{-18}$ and below~\cite{mcgrew_atomic_2018,bothwell_jila_2019,Brewer_IonClock_2019,additional_ref_1,additional_ref_3}, 
they are becoming a sensitive probe of the point to point geopotential at the sub-centimeter scale, where they are expected to %outperform
complement other methods of geodesy~\cite{geodesy_1,geodesy_2,geodesy_3,Delva_ProspectsGeo_2013,takano_geopotential_2016,lion_determinationGeo_2017,grotti_geodesy_2018,mcgrew_atomic_2018,additional_ref_2}. For example, a blinded comparison between two independent Yb optical lattice clocks was recently performed with accuracy, instability and reproducibility all at the level required to resolve sub-cm height differences~\cite{mcgrew_atomic_2018}. In addition, the frequency gradient due to the gravitational redshift across a single millimeter-scale atom ensemble was recently observed using Rabi spectroscopy of ${}^{87}$Sr without the use of a blinding offset and taking advantage of an 8-mHz linewidth clock laser~\cite{bothwell_resolving_2022}. 

Emerging clock applications such as relativistic geodesy require transportable optical clocks, which currently have poorer stabilities and accuracies than those of laboratory-based clocks \cite{ptb_transportable_2017,grotti_geodesy_2018,takano_geopotential_2016}. State-of-the-art laboratory-based optical clocks
often make use of bulky and immobile reference cavities with second-scale coherence times~\cite{kessler_SiCavity_2012,Zhang_SiCavity_2017,Robinson_SiCavity_2019} in order to achieve lower levels of clock instability and to aid in rapid systematic evaluation, limiting deployment in the field.
It has recently been demonstrated that differential measurements between single ions~\cite{Clements_IonCoherence_2020} and neutral atom ensembles~\cite{Takamoto_differential_2011,schioppo_ultrastable_2017,zheng_differential_2022},
as well as differential spectroscopy by phase-coherently linking a zero-dead-time optical lattice clock and a single ion clock~\cite{Kim_coherenceAtomicSpecies_2022},
allow interrogation times beyond the limit set by the local oscillators,
opening up the prospects for future applications with transportable or space-based clocks~\cite{derevianko_hunting_2014,kolkowitz_gravitational_2016,safronova_search_2018,Wcisło_dark_2018,kennedy_dark_2020}.

In this work, we perform a local-oscillator independent, blinded test of the gravitational redshift at the sub-centimeter scale using a spatially multiplexed optical lattice clock~\cite{zheng_differential_2022} consisting of an array of ${}^{87}$Sr atom ensembles trapped in a vertical, one-dimensional (1D) optical lattice (Fig.~1a).
We prepare 5 atomic ensembles equally spaced by 2.5 mm,
spanning a total height difference of 1 cm.
Synchronous differential comparisons are performed between the 5 ensembles,
resulting in 10 unique pairwise clock comparisons recorded simultaneously,
including 4 pairs at 2.5 mm,
3 pairs at 5.0 mm,
2 pairs at 7.5 mm,
and 1 pair at 1.00 cm height difference.
The gravitational redshift is tested by mapping out the frequency differences between each ensemble pair as a function of height difference.
We measure a fractional frequency gradient of $[-12.4\pm0.7_{\rm{(stat)}}\pm2.5_{\rm{(sys)}}]\times10^{-19}/$cm, consistent with the expected gravitational redshift gradient of $-10.9\times10^{-19}/$cm.
Our result constrains deviations from the redshift predicted by general relativity to $0.13\pm0.23$ for mm to cm scale height differences.

\begin{figure}[!ht]
    \centering
    \includegraphics[width=0.95\textwidth]{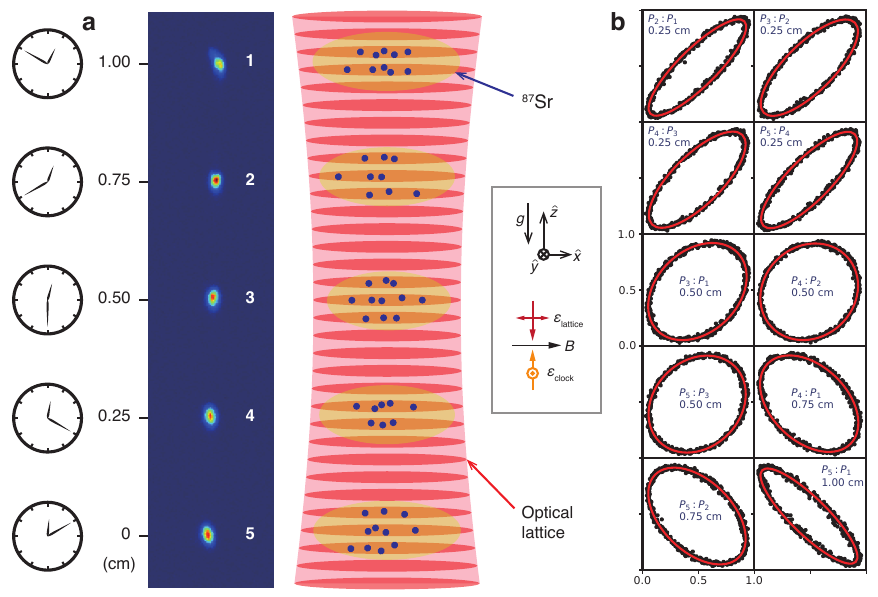}
    \caption{
    \textbf{$\vert$Experimental system and synchronous differential clock comparisons.}
    {\bf a}, A representative camera image of a spatially multiplexed array with 5 ensembles of ${}^{87}$Sr atoms (indexed 1--5 from top to bottom) trapped in a vertical 1D optical lattice for differential clock comparisons. The spacing between neighboring ensembles is 0.25 cm, spanning a total height difference of 1 cm. Due to the gravitational redshift, clocks at a higher gravitational potential are predicted to tick faster than clocks at a lower potential. The grey box shows the orientations of the applied bias magnetic field ($\bf{B}$), and the lattice and clock laser polarizations ($\epsilon$).
    {\bf b}, A representative outcome from synchronous Ramsey spectroscopy on the $\ket{{}^1S_0,m_F=+5/2}\leftrightarrow\ket{{}^3P_0,m_F=+3/2}$ clock transition with a 10~s free precession time using 5 atomic ensembles, resulting in 10 pairwise clock comparisons. In each plot, the excitation fractions of ensemble $j$ are plotted against the excitation fractions of ensemble $i$ ($P_j:P_i$, where $j>i$), tracing out an ellipse which is fitted to in order to extract the frequency difference between that pair. The frequency difference is dominated by the first-order differential Zeeman shift, which is rejected when averaging transitions between opposite spin states (Supplementary Note 3A).
    }
\end{figure}

\section*{Results}
\noindent \textbf{Testing the gravitational redshift with a miniature clock network}

The principles and basic operation of the multiplexed optical lattice clock used in this work were first described and experimentally demonstrated in Ref.~\cite{zheng_differential_2022}. In that work we demonstrated differential clock comparisons between atom ensembles with fractional frequency imprecision below $1\times10^{-19}$. However, there are a wide range of potential sources of differential frequency shifts at the mHz level, so the exact origin of the frequency differences we measured was not known, and the contributions from the relativistic gravitational redshift could not be independently extracted. In this work, we leverage several key improvements to our apparatus and experimental procedure that have been made since our prior work,
and perform a full systematic evaluation of all potential sources of differential frequency shifts at the $10^{-19}$ level, enabling a blinded test of the gravitational redshift with mm-scale sensitivity to gravitational potential differences.

Our major improvements include employing a deeper initial optical lattice ($130~E_{\rm rec}$ trap depth, where $E_{\rm rec}$ is the recoil energy of a lattice photon) during atomic ensemble loading and in-lattice cooling provide larger atom numbers ($>2000$ atoms per ensemble) with reduced atomic temperatures both axially ($>99\%$ occupancy in the lattice ground band) and radially ($<200$ nK).
Synchronized Ramsey measurements are subsequently performed at a shallower operational lattice depth ($u_{op}$) of $15~E_{\rm rec}$.
Combined with common-mode rejection of clock laser noise,
we achieve 32 s atom-atom coherence times,
more than 300 times longer than our measured atom-laser coherence time of roughly 100~ms~(Supplementary Note 2).
This enables differential instabilities below $1\times10^{-17}/\sqrt{\tau}$ for all the ensemble pairs,
a factor of 3 reduction in instability over our previous work~($3\times10^{-17}/\sqrt{\tau}$ with 6 ensembles).
In addition, we now suppress the residual magnetic field gradient along the lattice axis by a factor of more than 10,
reducing systematic uncertainties arising from Zeeman shifts.
These advances allow us to rapidly evaluate the differential systematic shifts at the $10^{-19}$ level due to environmental perturbations and thus perform a precision test of the gravitational redshift.
To the best of our knowledge, our work represents the first complete systematic evaluation of differential frequency shifts at the $10^{-19}$ level making use of synchronous Ramsey spectroscopy. This technique, which unlike differential Rabi spectroscopy can be used to probe well beyond the coherence time of the local oscillator, enables a new modality of precision measurement with optical lattice clocks where both the achievable accuracy and precision are now unbounded by the quality of the local oscillator.

For our measurements,
we utilize synchronous Ramsey spectroscopy in conjunction with spatially resolved fluorescence imaging to probe the clock transition along the ensemble array.
The optical lattice is operated near the magic wavelength where the differential polarizability between the ground (${}^1S_0, g$) and clock (${}^3P_0, e$) states is zero~\cite{brown_hyperpolarizability_2017,ushijima_operational_2018}.
We probe with an interleaved sequence using the magnetically insensitive $\ket{g, m_F = \pm 5/2}\leftrightarrow\ket{e, m_F = \pm3/2}$ clock transitions~\cite{Oelker_ClockStability_2019},
where $m_F$ is the projection of total angular momentum along the quantization axis defined by the bias magnetic field.
Taking the average of the clock transitions with opposite sign nuclear spin states rejects first-order Zeeman shifts and vector AC Stark shifts.
The differential phase $(\phi_d)$ for each ensemble pair is extracted through least squares ellipse fitting,
and is related to the differential frequency $(\delta f)$ for the pair through $\phi_d = 2\pi \delta f T_R$, where $T_R$ is the Ramsey free evolution time.
A representative outcome from clock interrogation on the $\ket{g,m_F=+5/2}\leftrightarrow\ket{e,m_F=+3/2}$ transition in shown in Fig.~1b,
where $(\delta f)$ is dominated by the first-order Zeeman shift (Supplementary Note 3A).

The differential frequency between each atomic ensemble pair includes a contribution from the gravitational redshift as well as other frequency shifts arising from differences between the two ensembles and their environments,
necessitating an evaluation of potential sources of systematic effects.
To avoid possible bias towards the expected result,
we adopt a blinded measurement protocol.
A blinded offset gradient was randomly drawn from a range of $\pm 5\times10^{-18}/$cm,
roughly 10 times the size of the expected redshift gradient,
and is automatically added to the extracted differential phase by our data analysis code following the ellipse fitting step.
This blinded offset was not known to the authors during systematic evaluation and data taking.
The blinded value of the measured frequency gradient across the array, the corrections for systematic shifts, and the systematic and statistical uncertainties were all finalized prior to the removal of the blind.

\noindent \textbf{Systematic effects and error budget}

\begin{figure}[!ht]
    \centering
    \includegraphics[width=0.95\textwidth]{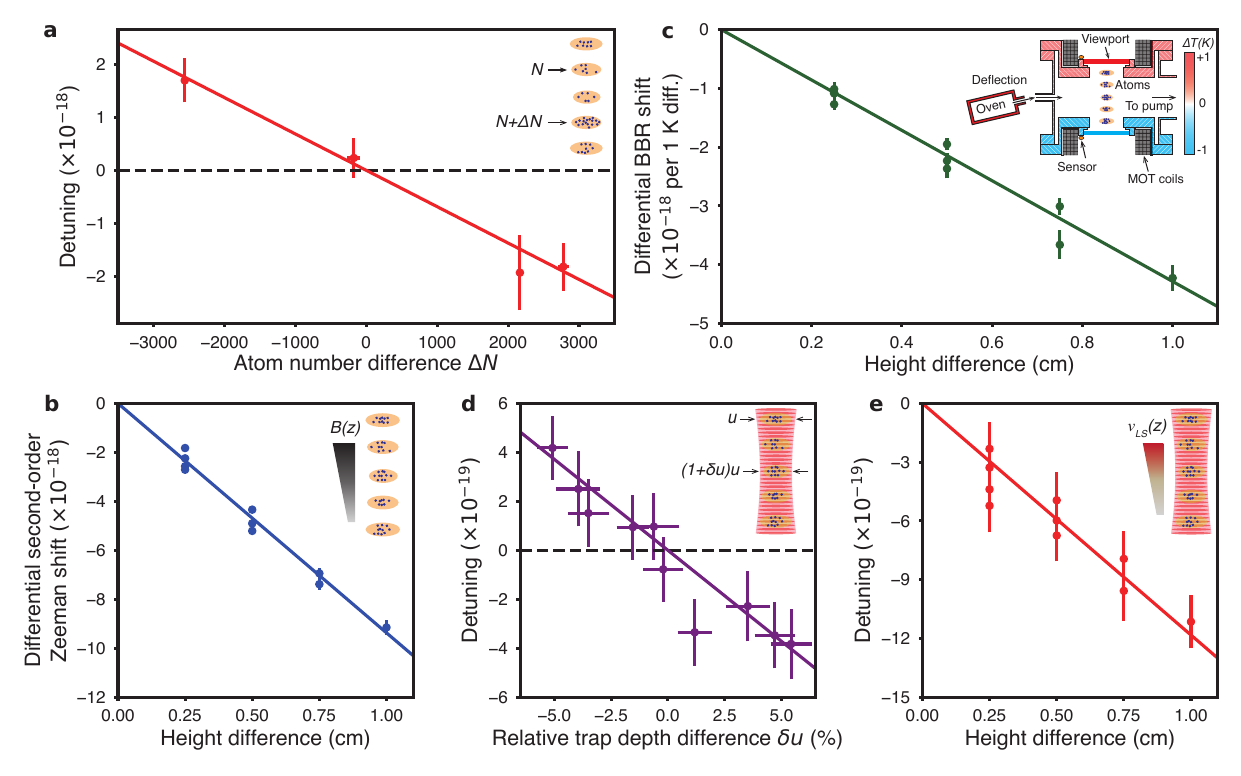}
    \caption{
    \textbf{$\vert$Sources and characterization of primary systematic shifts.}
    The error bars correspond to 1$\sigma$ standard deviation.
    {\bf a}, The differential density shift at $u_{op}=15~E_{\rm rec}$ for a symmetric pair (2, 4) is evaluated by varying the atom number difference ($\Delta N$). A linear fit yields a shift of $-0.7(1)\times10^{-19}$ per 100 atom number difference.
    {\bf b}, Evaluation of the second-order Zeeman gradient arising from the magnetic field gradient ($\partial B/\partial z\approx 1.5$ mG$/$cm). A linear fit yields a gradient of $-95.3(1.0)\times10^{-19}/$cm.
    {\bf c}, Characterization of the BBR shift due to thermal gradients across the vacuum chamber. The inset is an illustration of the science chamber. To evaluate the BBR effect, a thermal gradient is introduced by varying the temperature difference between top and bottom viewports by up to $\pm1$ K. A linear fit yields a BBR sensitivity of $-4.2(1)\times10^{-18}/$cm per 1 K difference in our system.
    {\bf d}, Correlations between relative trap depth difference ($\delta u$) and differential lattice light shifts after subtraction of the residual spatial light shift gradient. $u$ and $(1+\delta u)u$ correspond to the absolute trap depths for the ensemble pairs.
    {\bf e}, Evaluation of lattice light shift gradient at $u_{op}=15~E_{\rm rec}$ after removing contributions from $\delta u$ shifts. A linear fit yields a gradient of $-8.0(1.1)\times10^{-20}/E_{\rm rec}/$cm.
    }
\end{figure}

The results of the full systematic evaluation are listed in Table~1.
The procedure for measuring the contribution of each potential systematic is discussed in Supplementary Note 3.
Several effects dominate the extracted differential frequency and the corresponding systematic uncertainty, and we highlight them here.
First, atomic interactions due to $p$-wave collisions between on-site atoms lead to a frequency shift that scales linearly with atomic density~\cite{Martin_ManyBody_2013,Zhang_SUN_2014,Aeppli_DensityShift_2022}.
In our differential measurement,
the density shift is suppressed by a factor of roughly 10 compared to the absolute shift by balancing the number of atoms loaded into each ensemble.
By intentionally varying the atom number differences,
the differential density shift for a symmetric pair (2, 4) at $u_{op}$ is evaluated to be $-0.7(1)\times10^{-19}$ per 100 atom number difference (Fig.~2a).
Due to the Gaussian nature of the lattice beam,
we observe a minor trap volume dependence of the density shift,
which is accounted for in our evaluation by independently extracting the density shift as a function of atom number difference for each pairwise comparison (Supplementary Note 3B).
In each measurement run,
the corresponding differential density shifts are corrected for each ensemble pair individually.

The largest systematic shift in our system is the second-order Zeeman shift due to the background magnetic field gradient ($\sim1.5$ mG$/$cm).
The splitting of the transitions with opposite sign nuclear spin states provides a measurement of the magnetic field difference between each ensemble pair,
while the overall magnitude of the applied bias magnetic field ($\sim5.5$ G) is measured independently using a more magnetically sensitive transition.
This bounds the uncertainty from the second-order Zeeman gradient to below $1\times10^{-19}/$cm,
limited by uncertainty in the second order Zeeman coefficients for the clock transition (Fig.~2b and Supplementary Note 3A).

The frequency shift due to black body radiation (BBR) is the dominant source of systematic uncertainty for many room-temperature optical clocks\cite{SingleIon_-18_2016,mcgrew_atomic_2018,bothwell_jila_2019,Brewer_IonClock_2019}.
In our system, uncertainty arises due to 
temperature gradients in the surrounding environment. 
Because the ensembles are arrayed vertically, 
the primary contribution comes from differences in temperature between the top and bottom recessed viewports of the science chamber, which are the closest surfaces to the atoms.
To evaluate the effect of this BBR gradient,
we intentionally introduce a thermal gradient by varying the temperature difference between the two viewports by up to $\pm1$ K.
Mapping out the resulting differential frequency shifts of the 10 ensemble pairs,
we measure a linear BBR gradient sensitivity of $-4.2(1)\times10^{-18}/$cm per 1 K difference (Fig.~2c and Supplementary Note 3C).
This results in a weighted average BBR gradient of $-15.7(1.5)\times10^{-19}/$cm under normal operating conditions,
which is determined from the temperature differences recorded with calibrated temperature sensors that were continuously monitored during data taking.
This measurement also highlights the application of differential comparisons between spatially multiplexed ensembles for BBR gradient evaluation at the $10^{-19}/$cm level, which will be important for pushing the accuracy of room temperature optical lattice clocks below the $10^{-18}$ level. 

AC Stark shifts from lattice light contribute to the fractional frequency uncertainty of state-of-the-art optical lattice clocks at the low $10^{-18}$ level~\cite{brown_hyperpolarizability_2017,ushijima_operational_2018}.
For differential comparisons between ensembles within a single shared optical lattice,
this is significantly suppressed.
A differential lattice AC Stark shift between two ensembles is caused by the relative trap depth difference $\delta u$ arising from the lattice beam profile,
and scales linearly with $\delta u$ to first-order~(Supplementary Note 3D).
In conjunction with the multiplexed ensemble technique,
we can rapidly map out both $\delta u$ and the differential light shifts.
In doing so, we also observe a residual spatial light shift gradient of $-8.0(1.1)\times10^{-20}/E_{\rm rec}/$cm,
which depends linearly on the lattice depth and the spatial separation between ensembles, and does not depend on $\delta u$.
We believe this gradient is likely due to the differential tensor Stark shift arising from slight variations in the orientation of the magnetic field vector across the ensemble array. This is supported by the observation of a differential vector Stark gradient of $-2.5(2)\times10^{-18}/E_{\rm rec}/$cm in our system~(Supplementary Note 3D).
Regardless of the exact origin of the spatial light shift gradient,
we are able to measure and account for it by varying the lattice depth.
Subtracting off this residual spatial gradient,
we observe correlations between $\delta u$ and the remaining shifts~(Fig.~2d),
as expected.
This allows us to extract the operational lattice detuning from the effective magic wavelength~\cite{ushijima_operational_2018,bothwell_resolving_2022},
and to independently evaluate the lattice light shift gradient upon removal of the $\delta u$ shifts (Fig.~2e).

\noindent \textbf{Data analysis and interpretation}

We performed 14 blinded measurements of the gravitational redshift under normal operating conditions over a 3-week data taking campaign.
In each measurement run (ranging in duration from 1 to 4 hours),
the frequency gradient is determined by fitting a linear slope to the 10 measured differential frequencies as a function of the pairwise height differences,
after taking into account systematic corrections such as ellipse fitting bias and
density shift.
In order to account for correlations that arise between clock comparison pairs that share the same clock, the covariance between the pairwise comparisons is included in the error estimation (see Methods for details).
Corrections for spatially varying systematics such as the lattice light shift,
BBR shift,
and second-order Zeeman shift are applied to the measured gradient.
Upon removal of the blinded offset gradient, we find a weighted mean fractional frequency gradient of $[-12.4\pm 0.7_{(\rm stat)}\pm2.5_{(\rm sys)}]\times10^{-19}/$cm 
(where `stat' and `sys' indicate statistical and systematic uncertainties, respectively),
consistent with the expected redshift gradient of $-10.9\times10^{-19}/$cm within $1\sigma$ total uncertainty~(Fig. 3a).
The statistical uncertainty is scaled up by the square root of the reduced $\chi^2 $ statistic,
$\chi^2_{\rm red}=1.16$.
Our measurement is inconsistent with the hypothesis that there is no gravitational redshift on the surface of the Earth for mm-cm scale height differences at a confidence level of $4.9\sigma$.

\begin{figure}[!ht]
    \centering
    \includegraphics[width=0.95\textwidth]{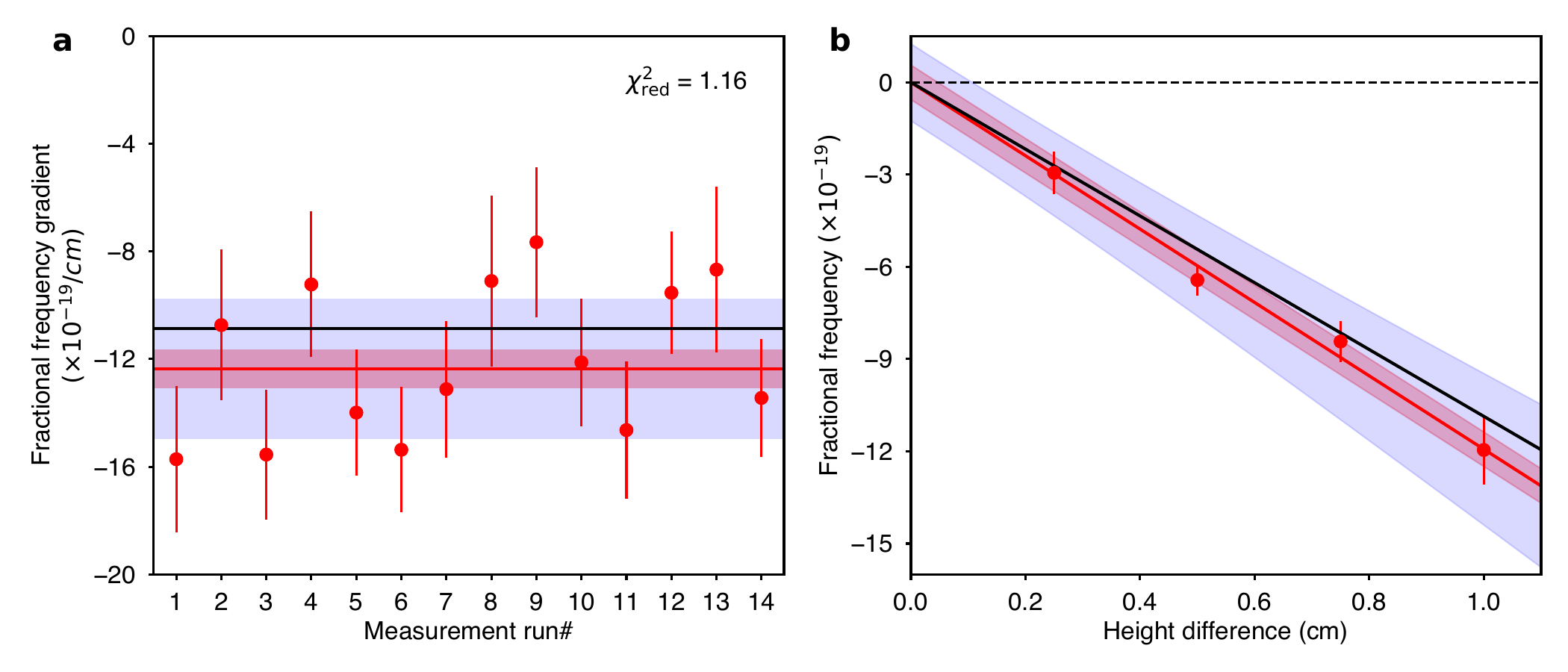}
    \caption{
    \textbf{$\vert$Gravitational redshift measurements.}
    {\bf a}, The measured fractional frequency gradients after accounting for all systematic corrections over 14 data taking runs (red scatter points), each of duration ranges from 1 to 4 hours. The weighted average (red solid line) yields a measured gradient of $[-12.4\pm0.7_{(\rm stat)}\pm2.5_{(\rm sys)}]\times10^{-19}/$cm, consistent with the expected redshift gradient (black solid line). Red (blue) area represents $\pm1\sigma$ statistical (total) uncertainty, in which the statistical uncertainty is inflated by the square root of the reduced $\chi^2 $ statistic, $\chi^2_{\rm red}=1.16$. The error bars correspond to 1$\sigma$ standard deviation.
    {\bf b}, Mean differential frequencies as a function of height difference across all measurements (red scatter points), analyzed using the same data set as in {\bf a}. A linear fit (red solid line) yields a fractional frequency gradient of $(-11.9\pm2.6)\times10^{-19}/$cm, again fully consistent with the expected redshift gradient (black solid line).
    }
\end{figure}

Deviations from the gravitational redshift predicted by general relativity can be parameterized by defining a modification parameter $\alpha$
\begin{equation}
    \frac{\delta f}{f} = (1+\alpha)\frac{g \Delta h}{c^2}
    \label{modified_rs}
\end{equation}
to first-order of the gravitational potential difference~\cite{Will_GR_2014,takamoto_test_2020}.
We constrain deviations from the predicted scaling by $\alpha=0.13\pm0.23$ for millimeter to centimeter scale height differences.
We note that while the most stringent constraints on $\alpha$ are at the $10^{-5}$ level \cite{Redshift_Eccentric_2018,Redshift_Eccentric2_2018}, those measurements were performed at very different length scales, with height differences roughly a factor of $10^{9}$ times larger than those used here.

In the future it appears feasible to extend our measurement approach to a larger apparatus in order to achieve spatial separations between ensembles on the scale of roughly 1 meter, a scale of laboratory atomic physics experiment that has already been demonstrated with atom interferometers~\cite{AI_Stanford_2015},
which would offer an increase in the magnitude of the redshift by a factor of $100\times$ over the separations used in this work.
In addition, we see a clear path to reducing the systematic uncertainty of our differential comparisons by more than one order of magnitude,
which can be accomplished by mitigating density shifts via operating at ``magic'' excitation fractions (as was recently demonstrated within a single ensemble \cite{bothwell_resolving_2022}) or through modifications to the lattice geometry \cite{campbell_fermi_degenerate_2017},
reducing the differential tensor lattice light shift by eliminating background magnetic field gradients, and mitigating the black-body radiation gradients through improved control of the thermal environment. When combined, such an experiment would promise constraints on $\alpha$ at the $10^{-4}$ level, indicating that laboratory-based tests of the gravitational redshift could soon be competitive with space-based tests \cite{Redshift_Eccentric_2018,Redshift_Eccentric2_2018} or tests with portable clocks \cite{takamoto_test_2020}.

As an alternative to the above analysis,
we reanalyze the same experimental data by taking the weighted average of the differential frequencies of each ensemble pair from the 14 measurement runs,
with systematic corrections applied individually to each pair.
The weighted mean differential frequencies of ensemble pairs with the same height differences (0.25, 0.50, 0.75 and 1.00 cm) are shown in Fig.~3b.
Through a final linear fit,
we find a frequency gradient of $(-11.9\pm2.5)\times10^{-19}/$cm,
again fully consistent with the expected redshift.
Extrapolating the spatially varying systematic uncertainties to 0 cm of separation,
we find our gravitational redshift resolution to be 1.3 mm, dominated by systematic uncertainty due to the differential density shift,
which could potentially be further reduced by incorporating the recent technique for density shift cancellation~\cite{Aeppli_DensityShift_2022} in future work~(Supplementary Note 3B).

\newpage

\begin{table}
\centering
\begin{tabular}
{ | l | c | c | }
 \hline
 Sources &  Gradient    & Uncertainty  \\
  &  ($\times10^{-19}/$cm)   &  ($\times10^{-19}/$cm) \\
 \hhline{|=|=|=|}
 BBR$^a$  &  -15.7   &  1.5 \\
 \hline
 Lattice light  &  -11.8   &  1.2 \\
 \hline
 Density$^b$  &   --  &  1.0 \\
 \hline
 $2^{\rm{nd}}$ order Zeeman$^a$  &   -95.3  &  1.0 \\
 \hline
  Probe Stark  &  0  &  0.5 \\
 \hline
  DC Stark  &  0  &  0.1 \\
 \hline
  Ellipse fitting$^b$ &  --  &  0.5 \\
 \hhline{|=|=|=|}
 Total systematic correction &  +122.8   &  2.5 \\
 \hline
 Statistical gradient  &  -135.2   &  0.7 \\
 \hhline{|=|=|=|}
 Corrected gradient  &  -12.4   & 2.6     \\
 \hline
 Expected redshift gradient   &  -10.9   & $<0.1$     \\
 \hline
\end{tabular}
\end{table}

\noindent {\bf Table 1. Fractional frequency gradients and corresponding uncertainties.} 
Uncertainties are quoted as 1$\sigma$ standard deviations.
For each systematic effect,
more discussion can be found in Supplementary Note 3.
\newline $^a$ The BBR shift and second-order Zeeman shift are corrected for each measurement,
here the weighted averaged values across all 14 measurements are listed.
\newline $^b$ The density shift and bias error from ellipse fitting are corrected for each pairwise clock comparison individually, and do not depend linearly on spatial separation.

\section*{Discussion}

\begin{figure}[!ht]
    \centering
    \includegraphics[width=0.95\textwidth]{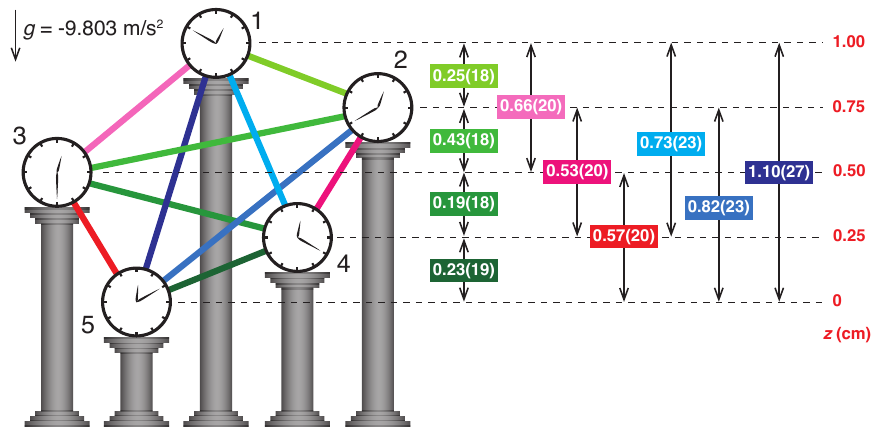}
    \caption{
    \textbf{$\vert$Extracting relative height differences using relativistic gravitational potential measurements.}
    The relative clock height differences across the array are determined using the measured gravitational redshifts and the independently measured local gravitational acceleration $g$. The double arrows represent the extracted height difference and the associated uncertainty for each clock pair. The true clock heights are shown on the right (red values), with the lowest clock (clock 5) defined as being at a height of 0 cm. All units are in cm unless otherwise specified.
    The uncertainties correspond to 1$\sigma$ standard deviation.
    }
\end{figure}

The measurement of gravitational potential differences between clocks with sub-centimeter resolution is a major goal of relativistic geodesy ~\cite{Delva_ProspectsGeo_2013,takano_geopotential_2016,lion_determinationGeo_2017,grotti_geodesy_2018,mcgrew_atomic_2018}. In the preceding discussion and analysis we treated the height differences between each ensemble pair and the local gravitational acceleration $g$ as known, as we measure them independently, but did not \emph{a priori} trust the gravitational redshift predicted by general relativity. From another perspective, our measurements can be viewed as a proof-of-principle demonstration of %relativistic geodesy
relativistic gravitational potential measurement
with millimeter scale resolution. Taking as a given that the redshift is given by Eq.~\ref{rs} and treating the ensemble array as a network of spatially distributed clocks with unknown height differences,
we can extract the height ordering and relative height differences from the measured gravitational redshifts and $g$, which we measure independently.
We find that we correctly assign the order of gravitational potential differences within the network, and that all of the extracted height differences are within 2 mm of the known values~(Fig.~4).
%beyond the geodetic limit
However, we note that we greatly benefit from the rejection of common-mode systematic shifts thanks to the ensembles sharing the same optical lattice and the same science chamber, which will not be possible when comparing two individual clocks at different geospatial locations. In addition, over a long baseline ($>1$ km), phase noise from the frequency transfer will not be common-mode and will limit the coherence times of the differential comparison.
Therefore, while these results demonstrate that relativistic gravitational potential measurements with mm-scale height resolution are achievable in the lab over short spatial separations, considerable challenges must be overcome before they can be applied to relativistic geodesy at length scales of interest.

In a recent work~\cite{bothwell_resolving_2022}, Bothwell and collaborators resolved the gravitational redshift across a single 1 mm atom ensemble.
While there are aspects in common between this work and Ref.~\cite{bothwell_resolving_2022}, there are also several critical differences that set this work apart. First, we employed a blinded offset during data taking and systematic evaluation. Second, while Bothwell et al.~made use of second-scale Rabi spectroscopy with an 8 mHz linewidth clock laser, we demonstrate comparable levels of differential stability and perform a full systematic evaluation at the $10^{-19}$ level by employing synchronous Ramsey spectroscopy with a Hz linewidth clock laser. This demonstrates that measurements of this kind need not be limited be the stability of the local oscillator. Third, we measure between spatially resolved ensembles using techniques that are likely more relevant to applications that require spatially separated clocks such as relativistic geodesy and gravitational wave detection. Finally, we also observe and characterize several systematic effects that were not observed by Bothwell et al., such as a black-body radiation gradient shift and differential tensor lattice light shift, likely in part because of the larger range of spatial separations used in our work. 

% conclusion should be removed
%\noindent \textbf{Conclusion} 

In conclusion, we perform a blinded, precision test of the gravitational redshift on the sub-centimeter scale with 5 spatially multiplexed ensembles of ${}^{87}$Sr.
We observe a gravitational redshift for millimeter to centimeter scale differences in height, and find that it is consistent with the expected general relativity gravitational redshift to within $1\sigma$ total uncertainty.
Our result is inconsistent with zero gravitational redshift at a 4.9 $\sigma$ confidence level and constrains deviations from the redshift predicted by general relativity to $0.13\pm0.23$ for mm to cm scale height differences.
We demonstrate a gravitational redshift measurement resolution of 1.3~mm.
Our results highlight the use of the spatially multiplexed ensemble techniques for achieving long coherence times and low differential instabilities without the need for a state-of-the-art clock laser,
and demonstrate its utility for characterization of spatially varying systematic shifts in optical lattice clocks on the sub-centimeter scale and at the $10^{-19}$ level.
These results represent an important milestone along the way to
%relativistic geodesy
gravitational potential measurements
at the sub-centimeter scale with optical atomic clocks~\cite{Delva_ProspectsGeo_2013,takano_geopotential_2016,lion_determinationGeo_2017,grotti_geodesy_2018,mcgrew_atomic_2018},
%the realization of entanglement enhanced clocks~\cite{Gil_spinsqueezing_2014,Pedrozo_entanglement_2020,VanDamme_squeezing_2021},
and explorations of the interplay between quantum mechanics and gravity~\cite{entanglement_gravity_1975,indirect_quantumgravity_1981,Zych_Proper_Time}.

\clearpage

\section*{\hfil Methods \hfil}

\subsection*{\hfil Sample preparation and experimental procedure\hfil}

The experimental sequence starts with laser cooling the atoms down to $1~\mu$K temperature with standard two-stage magneto-optical trapping (MOT).
Using the multiplexed ensemble loading technique with a movable one-dimensional (1D) optical lattice described and demonstrated in Ref.~\cite{zheng_differential_2022},
5 ensembles of ultra-cold, spin-mixed ${}^{87}$Sr atoms are loaded into a vertical optical lattice with a depth of $130~E_{\textrm{rec}}$,
where $E_{\textrm{rec}}/h \approx 3.5$ kHz is the recoil energy of a lattice photon and $h$ is the Planck constant,
with an equal spacing between ensembles of 0.25 cm over a total extent of 1 cm vertically.
This is followed by hyperfine spin polarization into either stretched state ($\ket{{}^1S_0, m_F=\pm9/2}$) and in-lattice cooling~(Supplementary Note 1).
The lattice is then adiabatically ramped down to the operational trap depth ($u_{op} =15~ E_{\textrm{rec}}$),
at which a series of $\pi$ pulses addressing the ${}^1S_0\leftrightarrow {}^3P_0$ ($g\leftrightarrow e$) transitions prepare the atoms into $\ket{e, m_F=\pm3/2}$.
Ramsey spectroscopy is performed by interrogating the $\ket{g, m_F=\pm5/2}\leftrightarrow\ket{e, m_F=\pm3/2}$ clock transitions.
Following the second Ramsey $\pi/2$ pulse, the lattice is adiabatically ramped back up to 130 $E_{\textrm{rec}}$ for read-out.
The populations in the ground and excited clock states of all 5 ensembles are read-out in parallel with imaging pulses along the lattice axis,
with scattered photons collected on a camera (Andor, iXon-888).
The excitation fraction is extracted through
$P = (N_e-N_{bg})/(N_g+N_e-2N_{bg})$,
where $N_g, N_e$ and $N_{bg}$ are the ground state population,
excited clock state population, and background counts without atoms, respectively.

As described in Ref.~\cite{zheng_differential_2022}, for multiple ensemble preparation,
we chirp the frequency of the retro-reflected lattice beam during the single-frequency stage of the narrow-line ${}^1S_0\leftrightarrow{}^3P_1$ second-stage MOT.
We perform 4 lattice movements of 0.25 cm each in order to load 5 spatially separated atom ensembles,
and a final lattice movement of 0.5 cm in the opposite direction,
which positions the ensemble array symmetrically around the lattice beam waist.
The entire duration of the moving lattice portion of the loading sequence is typically less than 100 ms,
with 80\% of total atom number transferred efficiently from the narrow-line MOT to the 5 ensembles.
In-lattice axial (sideband) and radial (Doppler) cooling are subsequently applied to lower the atom temperatures after lattice acceleration.

For clock interrogation,
we probe the $\ket{g, m_F = \pm5/2}\leftrightarrow \ket{e, m_F = \pm3/2}$ transitions with a shared clock laser along the lattice axis.
Synchronous Ramsey spectroscopy is performed to reject common-mode local oscillator noise.
The typical Ramsey interrogation time ($T_R$) is roughly 10 s, 
with a dead time ($T_d$) of 2 s between interrogations for sample preparation and read-out,
yielding a measurement duty cycle of 83\%.
Simultaneously probing 5 ensembles results in 10 pairwise clock comparisons for a single nuclear spin state.
Combined with atom numbers ($N$) of $\sim 2000$ per ensemble and contrast ($C$) of above 80~\%,
the typical differential instability for each pairwise comparison is below $1\times10^{-17}/\sqrt{\tau}$,
where $\tau$ is the averaging time,
consistent with the quantum projection noise (QPN) limit
\begin{equation}
    \sigma_{\rm QPN}(\tau) = \frac{\sqrt{2}}{2\pi f C T_R}\sqrt{\frac{T_R+T_d}{N\tau}},
\end{equation}
where $f$ is the clock frequency,
$\tau$ is the averaging time,
and the factor of $\sqrt{2}$ assumes equal contribution from each clock.

\subsection*{\hfil Ellipse phase extraction \hfil }

We perform synchronous Ramsey spectroscopy with up to 5 ensembles
(indexing $1, 2,\dots, 5$ from top to bottom).
This results in 10 pairs of clock comparisons performed simultaneously when probing transition from either nuclear spin states $\ket{g,m_F=\pm5/2}$.
For each pair of clock comparison,
we plot the excitation fractions $P_i(P_j)$ of ensemble $i(j)$ on the x(y)-axis
(note that we choose the convention $j>i$).
The excitation fractions are given by
\begin{equation}
    \begin{split}
        P_i &= \frac{1}{2}\bigg[1+C_i{\rm cos}(\theta_L )\bigg],\\
        P_j &= \frac{1}{2}\bigg[1+C_j{\rm cos}(\theta_L + \phi_{d})\bigg],
    \end{split}
\label{excfrac_ellipse}
\end{equation}
where $C_{i(j)}$ is the contrast of ensemble $i(j)$,
$\theta_L$ is the common-mode laser phase,
and $\phi_d$ is the differential phase which yields the differential frequency ($\delta f_{ij}=f_j-f_i$) between ensemble pair $(i,j)$ through $\phi_d=2\pi\delta f_{ij} T_R$ for a given known Ramsey free evolution time $T_R$.

Since we are operating at Ramsey dark times well beyond the laser coherence times, $\theta_L$ is random and uniformly distributed from $0$ to $2\pi$.
The data randomly samples from points lying on an ellipse (with slight deviations from the ellipse due to QPN).
We then fit to this ellipse using a least-squares approach~\cite{ellipse_fitting}.
To extract the differential phase~\cite{estey_thesis},
we rewrite the data $\{P_i, P_j\}$ (denoted as $\{x, y\}$ below)
in the form of a generalized conic section
\begin{equation}
    a_1x^2 + a_2xy + a_3y^2 + a_4x +a_5y +a_6 = 0,
\label{ellipse_ai}
\end{equation}
which describes an ellipse when $a_2^2 -4a_1a_3<0$.
We rewrite Eq.~\ref{excfrac_ellipse} as
\begin{equation}
    \begin{split}
        x'&=\frac{2x-1}{C_x} = {\rm cos}(\theta_L),\\
        y'&=\frac{2y-1}{C_y} = {\rm cos}(\theta_L+\phi_d).
    \end{split}
\end{equation}
Through cancelling out $\theta_L$,
we have
\begin{equation}
    \begin{split}
        &\frac{4}{C_x^2}x^2 - \frac{8{\rm cos}\phi_d}{C_x C_y} + \frac{4}{C_y^2}y^2 + \bigg(\frac{4}{C_x C_y} - \frac{4}{C_x^2}\bigg)x + \bigg(\frac{4}{C_x C_y} - \frac{4}{C_y^2}\bigg)y\\
        & +\bigg(
        \frac{1}{C_x^2} - \frac{2}{C_x C_y} + \frac{1}{C_y^2} - {\rm sin}^2\phi_d
        \bigg) = 0,
    \end{split}
\end{equation}
which can be matched up with the coefficients $a_i$ from Eq.~\ref{ellipse_ai}.
The differential phase $\phi_d$ is then extracted using:
\begin{equation}
    \phi_d = {\rm cos}^{-1}\bigg(\frac{-a_2}{2\sqrt{a_1a_3}}\bigg).
\end{equation}

The associated Allan deviation is extracted via jackknifing technique~\cite{marti_prl_2018},
and is then fitted to a white frequency noise model with $1/\sqrt{\tau}$ scaling.
Extrapolating the fit to the full averaging time yields the statistical uncertainty of the differential frequency.

In our measurements,
we probe with an interleaved sequence between clock transitions with either nuclear spin state,
$\ket{g, m_F=\pm5/2}\leftrightarrow\ket{e, m_F=\pm3/2}$.
This results in 10 ellipses for transition with a single nuclear spin state,
and thus 20 ellipses per measurement.
A representative plot of the $\ket{g, m_F=+5/2}\leftrightarrow\ket{e, m_F=+3/2}$ transition is shown in Fig.~1b.
The differential phase for each ellipse is dominated by the differential first-order Zeeman shift (Supplementary Note 3A),
which is on the order of $\pm8\times10^{-17}/$cm and is rejected by averaging transitions with opposite spin states.

\subsection*{\hfil Data blinding protocol \hfil}

To eliminate possible bias of our data taking and systematic analysis towards an expected outcome,
we employ a data blinding protocol. Our data software adds a large constant offset gradient to our measurements,
including the data taken for systematic evaluations and data runs taken under normal operating conditions.
The blinded offset gradient is pseudo-randomly drawn from a uniform distribution spanning over $\pm5\times10^{-18}/$cm,
10 times the size of the expected redshift gradient.
The blinded offset gradient is scaled by the height difference between each ensemble pair,
and is then automatically added to the results from our data analysis code for ellipse phase extraction.

The blinded offset gradient was only unblinded after finalizing the corrections for all systematic effects,
determining the measured value with blinded offset gradient taken under normal operating conditions,
and finalizing the associated statistical and systematic uncertainties.
No additional data was taken and no changes were made to the analysis, the error budget, the measured value, or the uncertainties after unblinding.

\subsection*{\hfil Normal operation, data taking, unblinding and analysis \hfil}

We performed 14 blinded measurement runs of gravitational redshift data under normal operating conditions over a 3-week campaign.
Each run ranged in duration from 1 to 4 hours,
and was performed in conjunction with verification of several experimental parameters to ensure that the associated systematic effects are under control,
such as the magnetic field gradient,
density shift coefficients,
$\delta u$,
and clock and lattice beam alignments (See Methods).
In each measurement run,
the differential frequency of each ensemble pair was extracted through ellipse fitting,
with the associated Allan deviation extrapolated to the full averaging time taken as the statistical uncertainty.
The corrections for density shifts and bias error from the ellipse fitting are applied to each ensemble pair individually.
The total uncertainty of each clock comparison is given by the quadrature sum of its statistical uncertainty and the uncertainties of systematic corrections.
We analyze the measured frequency gradient using two approaches.

In the first approach,
the extracted frequency differences for 10 ensemble pairs from each measurement run are plotted as a function of the height differences.
A linear fit is applied to each measurement run.
Many of the clock comparison pairs share a clock, e.g., pair (1, 2) and pair (2, 3) share clock 2.
This means that the quantum projection noise is partially correlated between pairs, and not accounting for this would result in an underestimation of the error bar associated with the fit.
To account for this,
the covariance matrix is included in the fitting algorithm,
where the covariance between clock pairs (a, b) and (b, c) is given by the jackknifing re-sampling approach~\cite{marti_prl_2018}
\begin{equation}
    \textrm{Cov}(\phi) = \frac{N-1}{N} \sum_i \bigg(\bar\phi_{ab}^{JK} - \bar\phi_{ab, \neq i}^{JK} \bigg) \bigg(\bar\phi_{bc}^{JK} - \bar\phi_{bc, \neq i}^{JK} \bigg),
\label{cov}
\end{equation}
where $\bar\phi^{JK}_{\neq i}$ is the extracted phase except the $i^{\rm th}$ data point,
$\bar\phi^{JK}$ is the mean of $\bar\phi^{JK}_{\neq i}$,
and $N$ is the total number of measurements.

The associated fitted slopes from 14 measurement runs,
after accounting for systematic gradient corrections,
are then weighted averaged, yielding a statistical uncertainty of $0.7\times10^{-19}/$cm,
inflated by the square root of the reduced $\chi^2 $ statistic,
$\chi^2_{\rm red}=1.16$
(Fig.~3a).
Upon completion of the measurements, the pseudo-randomly generated offset blinding gradient was revealed and subtracted from the measurements.
The offset gradient proved to be $+3.7\times10^{-18}/$cm,
and the measurements before and after unblinding are shown in Supplementary Figure~8.
We find a weighted mean frequency gradient of $[-12.4\pm 0.7_{(\rm stat)}\pm2.5_{(\rm sys)}]\times10^{-19}/$cm,
consistent with the expected redshift gradient of $-10.9\times10^{-19}/$cm within 1$\sigma$ total uncertainty.

In the second approach,
we re-analyze the data and apply systematic corrections to each pairwise clock comparison individually over the same raw data set.
For each pair,
the total uncertainty is calculated as the quadrature sum of the standard error of its weighted mean,
systematic uncertainties that don't scale with height difference (density shift and ellipse fitting corrections),
and other systematic uncertainties that scale with height difference.
The weighted averaged frequency differences of each ensemble pair are given in Supplementary Table~1.
Through a final linear fit to the differential frequencies as a function of the height differences,
we find a frequency gradient of $(-11.9\pm2.5)\times10^{-19}/$cm (Fig.~3b),
again fully consistent with the expected redshift gradient within $1\sigma$ total uncertainty.

\subsection*{\hfil Relativistic clock height difference measurements \hfil}

Assuming that theory of general relativity is correct and that the gravitational redshift is given by Eq.~\ref{rs},
and treating the ensemble array as a network of spatially distributed clocks with unknown heights,
we demonstrate relativistic gravitational potential measurement \cite{takano_geopotential_2016,grotti_geodesy_2018,mcgrew_atomic_2018} using synchronous clock comparisons (Supplementary Table~1).
The height difference $\Delta h$ for each clock pair is then given by
\begin{equation}
    \Delta h = \frac{\delta f}{f} \frac{c^2}{g},
\end{equation}
where $\delta f$ is the measured gravitational redshift,
$g$ is the independently measured local gravitational acceleration ($g=-9.803$ m$/$s$^2$),
$f$ is the clock transition frequency, and
$c$ is the speed of light.
The uncertainties of the extracted height differences are dominated by the systematic uncertainties of the measured gravitational redshifts.

\subsection*{\hfil Data availability \hfil}
The process data used in this study have been deposited on Zenodo digital repository under https://doi.org/10.5281/zenodo.8184043.

\subsection*{\hfil Code availability \hfil} 
The code used for experimental control, data analysis, and simulation in this work are available from the corresponding author upon reasonable request.

\subsection*{\hfil Acknowledgements \hfil} We acknowledge G.E.W.~Marti, A.~Jayich and T.~Bothwell for fruitful discussions and insightful comments on the manuscript. We acknowledge technical contributions from B.N.~Merriman, H.~Li, V.~Lochab and N.~Ranabhat. We are particularly grateful to B.~Tikoff, E. M.~Nelson and C.~Ruggles from the Department of Geoscience at the University of Wisconsin-Madison for performing measurements of the gravitational acceleration in our laboratory. 
This work was supported by the NIST Precision Measurement Grants program, the Northwestern University Center for Fundamental Physics and the John Templeton Foundation through a Fundamental Physics grant, the Wisconsin Alumni Research Foundation, a Packard Fellowship for Science and Engineering, a Sloan Research Fellowship, the Army Research Office through agreement number W911NF-21-1-0012, and the National Science Foundation under Grant No.~2143870.

\subsection*{\hfil Author contributions \hfil} All authors contributed to carrying out the experiments, data analysis and writing the manuscript.

\subsection*{\hfil Competing interests \hfil}
The authors declare no competing interests.

%\subsection*{\hfil Correspondence and requests for materials \hfil}
%Correspondence and requests for materials should be addressed to Shimon Kolkowitz.

\newpage

\section*{Supplementary Information for ``A lab-based test of the gravitational redshift with a miniature clock network"}

Xin Zheng$^{1,\dagger}$, Jonathan Dolde$^{1,\dagger}$, Matthew C. Cambria$^{1}$, Hong Ming Lim$^{1}$, 
Shimon Kolkowitz$^{1,2,\ast}$

\textit{\normalsize${}^{1}$Department of Physics, University of Wisconsin-Madison, WI 53706, USA}

\textit{\normalsize${}^{2}$Department of Physics, University of California, Berkeley, CA 94720, USA}

\textit{\normalsize${}^{\dagger}$These authors contributed equally to this work}

\textit{\normalsize${}^{\ast}$To whom correspondence should be addressed; E-mail: kolkowitz@berkeley.edu}

\section*{SUPPLEMENTARY NOTE 1: ATOMIC TEMPERATURES}

Representative data from clock sideband thermometry on the $\ket{g}\leftrightarrow \ket{e}$ transition at $130~E_{\textrm{rec}}$ is shown in Supplementary Figure~1a.
The reduced height of the red sideband ($n_z \rightarrow n_z-1$, where $n_z$ is the axial vibrational quantum number) indicates that $>99\%$ of the atoms are populated in the lattice ground band ($n_z=0$).
A separate clock probe perpendicular to the lattice axis taken at 15~$E_{\textrm{rec}}$ is shown in Supplementary Figure~1b,
in which the Doppler broadened profile indicates a radial atomic temperature below $200$ nK.

\captionsetup[figure]{labelfont={},name={Supplementary Figure}, labelsep=period}
 
\renewcommand{\thefigure}{1}
\begin{figure}[!ht]
    \centering
    \includegraphics[width=0.95\textwidth]{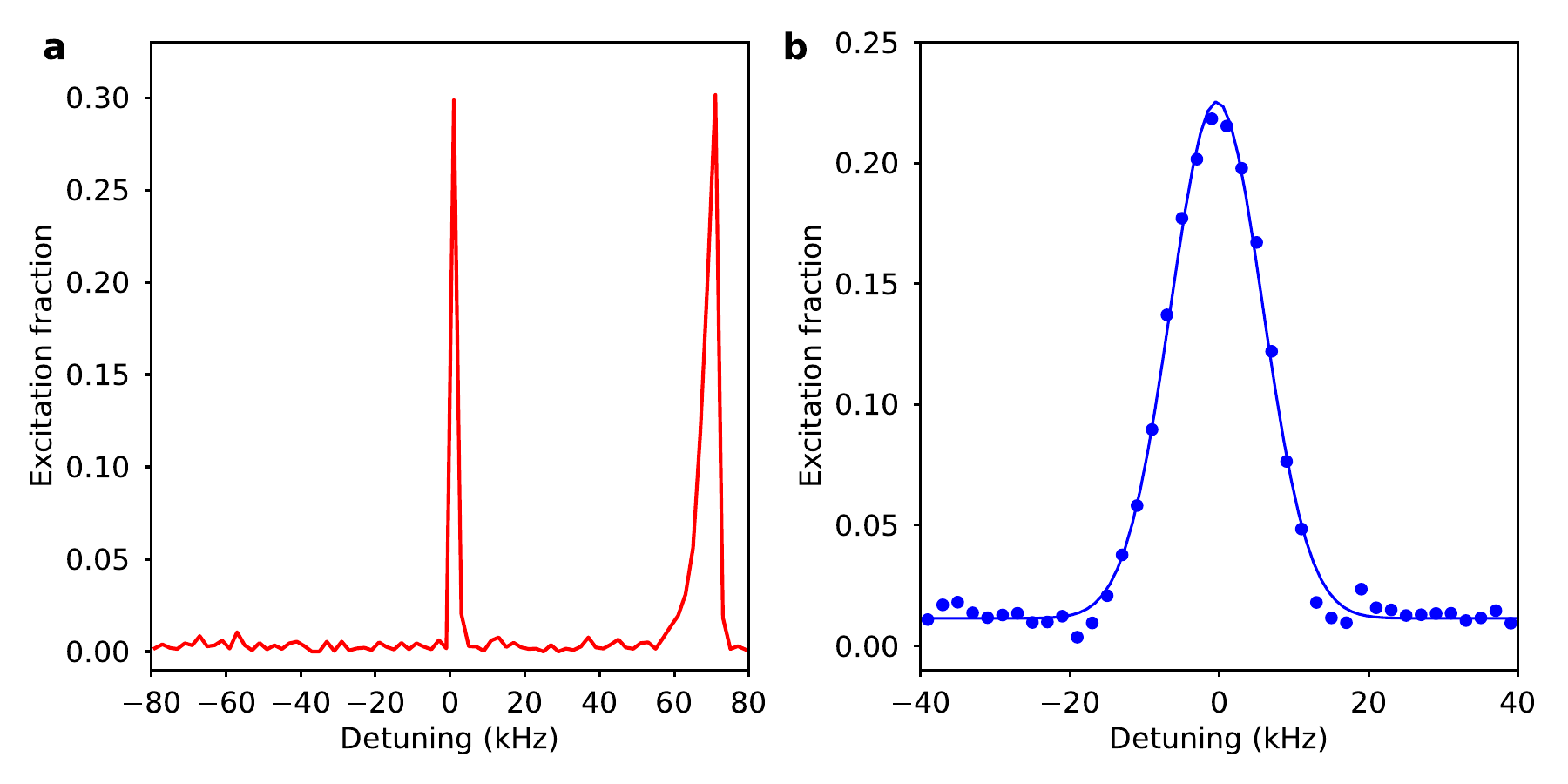}
    \caption{
    Atomic temperatures.
    (a) Axial clock sideband thermometry taken at 130 $E_{\rm rec}$ lattice trap depth following in-lattice cooling.
    The reduced height of the red sideband ($n_z\rightarrow n_z-1$) indicates that $99\%$ of atoms are populated in the lattice ground band ($n_z=0$).
    (b) Radial Doppler broadened profile taken at $15~{E_{\rm rec}}$. 
    A Gaussian fit (solid blue line) yields a radial temperature of 180 nK.
    }
\end{figure}

\section*{SUPPLEMENTARY NOTE 2: ATOMIC COHERENCE TIMES}

We use a synchronized Ramsey sequence to measure the atomic coherence.
We prepare two atomic ensembles in the $\ket{e, m_F=+3/2}$ state and apply a $\pi$/2 pulse on the $\ket{e, m_F=+3/2}\leftrightarrow \ket{g, m_F=+5/2}$ transition.
After waiting for a varying dark time ranging from 100 ms to 40 s,
we apply a second $\pi$/2 pulse with a random phase relative to the first.
We then measure the excitation fractions of each ensemble and plot the averaged contrast as a function of dark times.
We observe an exponential decay of the contrast with a time constant of $32.1(9)$ s at $15~E_{\rm rec}$ (Supplementary Figure~2),
300 times longer than the measured atom-laser coherence time ($\sim100$ ms).
We note that while the atomic coherence time is about 30 s,
the typical clock interrogation time in our measurements is around 10 s,
chosen to minimize the QPN because lost of atoms and reduced contrast at longer dark times~\cite{zheng_differential_2022s}.

\captionsetup[figure]{labelfont={},name={Supplementary Figure}, labelsep=period}
 
\renewcommand{\thefigure}{2}
\begin{figure}[!ht]
    \centering
    \includegraphics[width=0.70\textwidth]{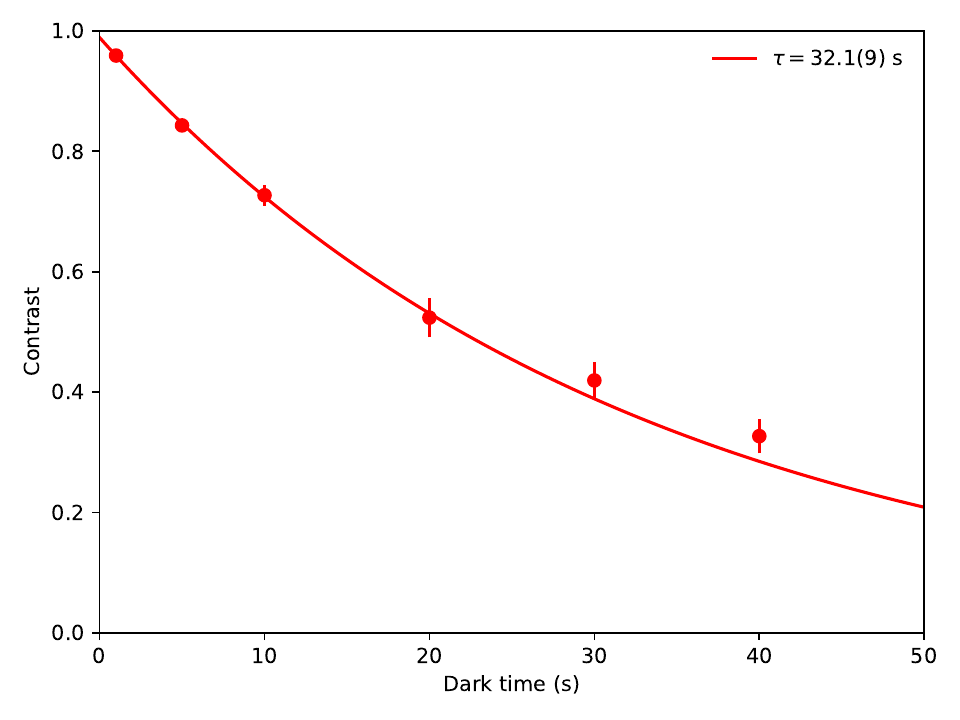}
    \caption{
    Atomic coherence times.
    Decay of Ramsey contrast as a function of dark times at $u_{op}=15~E_{\rm rec}$ (scatter points).
    An exponential fit (solid line) yields an atomic coherence time of 32.1(9) s.
    }
\end{figure}

\section*{SUPPLEMENTARY NOTE 3: SYSTEMATIC EVALUATION}

\subsection{Zeeman shifts}

Coupling of the clock states to external magnetic fields give rise to Zeeman shifts of the clock transition frequency.
For a single atomic ensemble with magnetic field amplitude $B$,
the Zeeman shifts when probing the $\ket{g, m_F = \pm 5/2}\leftrightarrow\ket{e, m_F = \pm3/2}$ clock transitions can be expressed as
\begin{equation}
    \nu_{\rm ZS}(B, \pm) = \pm\mu_L B +\mu_Q B^2,
\end{equation}
where $\mu_L$ and $\mu_Q$ are the first and second-order Zeeman shift coefficients, respectively.
For differential clock comparison between ensemble pair ($i, j$), where $j>i$,
with differing magnetic field strengths $B_i$ and $B_j$,
the differential Zeeman shifts,
$\Delta \nu_{\rm ZS}(B_i, B_j, \pm) = \nu_{\rm ZS}(B_j,\pm) - \nu_{\rm ZS}(B_i, \pm)$,
are given by
\begin{equation}
\begin{split}
    \Delta \nu_{\rm ZS}(B_i, B_j, \pm) &= \pm\mu_L (B_j - B_i) +\mu_Q (B_j^2 - B_i^2),\\
    &=\pm\mu_L \delta B + 2\mu_Q B \delta B, 
\end{split}
\end{equation}
where $B = (B_i + B_j)/2$ is the mean magnetic field amplitude,
and $\delta B = B_j - B_i$ is the magnetic field difference.
The first-order differential Zeeman shift (on the order of $\pm8\times10^{-17}/$cm) is rejected by averaging opposite spin state transitions,
while their splitting yields $\delta B$.
In the limit of $|\delta B| \ll B$,
the second-order differential Zeeman shift can be approximated as
\begin{equation}
\begin{split}
    \delta B &= \Delta\nu_{\rm split}/(2\mu_L),\\
    \Delta \nu_{\rm ZS, 2} &= \xi \Delta \nu_{\rm split} B,
\end{split}
\end{equation}
where $\xi = \mu_Q/\mu_L$,
and $\Delta \nu_{\rm split}$ is the splitting between the transitions with opposite spin states after subtraction of residual vector AC Stark shift.

The second-order Zeeman shift correction is applied for each measurement run,
with uncertainty primarily limited by $\xi$,
which is found to be $0.0105(1)$ G$^{-1}$ by varying $B$ and $\delta B$.
This is in good agreement with the theoretical value of $0.0104(2)$, calculated using $\mu_L=-22.4~$Hz$/$G and $\mu_Q = -0.233(5)$~Hz$/$G$^2$~\cite{boyd_pra_2007s}.
The bias magnetic field $B$ is calibrated each measurement day using the magnetically sensitive transitions $\ket{{}^1S_0, m_F=\pm9/2}\rightarrow\ket{{}^3P_1, m_F=\pm11/2}$,
and has a typical value of $\sim5.5$ G with fractional uncertainty below $10^{-4}$.
The field difference $\delta B$ ($\sim1.5$ mG$/$cm) is extracted from the splitting between the clock transitions with opposite nuclear spin states with fractional uncertainty below $10^{-3}$ after subtraction of the vector AC Stark shift.
Through weighted averaging of the data taken at normal operating conditions,
we find the mean second-order Zeeman gradient to be $-95.3(1.0)\times10^{-19}/$cm~(Fig.~2b in the main text).

\subsection{Density shift}

Due to the Pauli-exclusion principle,
$s$-wave interactions are forbidden for identical Fermionic atoms within a single lattice site,
while $p$-wave collisions are allowed,
leading to a clock frequency shift that scales linearly with atomic density~\cite{Martin_ManyBody_2013s,Zhang_SUN_2014s}.
In our system,
the differential density shift is evaluated by varying the atom numbers loaded into each ensemble for each individual pairwise comparison.
For a symmetric pair (2, 4),
we find a linear slope of $-0.7(1)\times10^{-19}$ per 100 atom number difference at our operational lattice trap depth $u_{op}=15~E_{\rm rec}$ (Fig.~2a in the main text).
In addition, we found a weak trap volume dependence due to the Gaussian nature of the lattice beam (less than 15~\% across the entire array),
which is accounted for in our evaluation as a different dependence on atom number difference for each pairwise comparison.
For a typical run of the experiment,
each atomic ensemble has about 2000 atoms,
corresponding to an overall density shift of about $-14(2)\times10^{-19}$.
The differential density shift is suppressed by a factor of 10 when the atom number difference is bounded below 200 through optimization of ensemble loading times and conditions.
In addition, we measure the atom number in each ensemble in every shot of the experiment,
and the average differential density shift is calculated and corrected individually for each pairwise clock comparison measurement run.
Uncertainty of density shift correction arises from shot-to-shot atom number fluctuations,
for which we estimate an upper bound limit of $1\times10^{-19}$.

A recent study found that operation in a gravity-tilted shallow lattice allows for cancellation of density shifts at a ``magic'' lattice depth near $12~E_{\rm rec}$, where the partially delocalized Wannier-Stark states enable tunability of on-site $p$-wave versus neighbouring-site $s$-wave atomic interactions~\cite{Aeppli_DensityShift_2022s}. In this work we did not observe such a cancellation effect at shallower lattice depths, likely due to the difference in dynamics between Rabi spectroscopy as was employed in Ref.~\cite{Aeppli_DensityShift_2022s} versus our use of Ramsey spectroscopy with a 50:50 superposition~\cite{boyd_thesis_s}. This offers the prospect of further reducing uncertainty from differential density shifts in future works.

\subsection{Black body radiation shift}

In our experiment,
the optical lattice is orientated nearly vertically,
with a tilt of about $5^\circ$,
and is centered with respect to the science chamber to the best of abilities.
The tilt is determined by measuring the Wannier-Stark ladder resonances (see the ``Wannier-Stark ladder and the expected gravitational redshift'' section below for details).
The recessed high-emissivity Fused Silica viewports (MPF Products) mounted on the top and bottom sides of the science chamber are the closest surfaces to the atoms, and are primarily responsible for the black body radiation (BBR) gradient along the lattice axis.
To study the BBR effect~\cite{beloy_BBR_2014s,ushijima_cryogenic_2015s},
we heat up either the top or bottom stainless steel flange of the science chamber,
which results in a temperature difference of up to $\pm1$ K between the top and bottom viewports (see inset of Fig.~2c in the main text.

The BBR shift due to thermal gradients between the top and bottom viewports can be expressed as~\cite{Haslinger_BBR_2018s}:
\begin{equation}
    \nu_{\rm BBR}(z) = -2\frac{\alpha_{\rm Sr}\sigma_{\rm SB}}{h\epsilon_0 c}\sum_{k={\rm top, bot}}\frac{\Omega_k(z)}{4\pi} T^4_k,
\end{equation}
where $h$ is the Planck constant,
$\alpha_{\rm Sr}$ is the atom's DC polarizability,
$\sigma_{\rm SB}$ is the Stefan-Boltzmann constant,
$\Omega_{k}(z)$ corresponds to the solid angle as seen by the atom ensemble at $z$,
and $T_k$ denotes the temperature on the surface of the viewport.
The differential BBR shift between an ensemble pair at $(z_i, z_j)$ is then given by
\begin{equation}
    \Delta\nu_{\rm BBR}(z_j, z_i) = -2\frac{\alpha_{\rm Sr}\sigma_{\rm SB}}{h\epsilon_0 c}
    \sum_{k={\rm top, bot}}\bigg(
    \frac{\Omega_k(z_j)}{4\pi} 
    -\frac{\Omega_k(z_i)}{4\pi} \bigg)
    T^4_k.
\end{equation}
Because the viewport separation ($2Z_0=15$ cm) is much larger than the separation between the ensemble pairs ($\delta z_{ji} = z_j-z_i \leq 1$ cm),
we approximate the solid angle difference as
\begin{equation}
\begin{split}
\Delta\Omega_{ji, \rm{top}} &= \pi r^2 \bigg(
\frac{1}{(Z_0+z_j)^2} - \frac{1}{(Z_0+z_i)^2}
\bigg) \approx -\frac{\pi r^2}{Z^3_0} \delta z_{ji},\\
\Delta\Omega_{ji, \rm{bot}} &= \pi r^2 \bigg(
\frac{1}{(Z_0-z_j)^2} - \frac{1}{(Z_0-z_i)^2}
\bigg) \approx \frac{\pi r^2}{Z^3_0} \delta z_{ji},
\end{split}
\end{equation}
where $r = 5.7$ cm is the radius of the viewport. 
The differential BBR shift can be further simplified through a Taylor expansion up to order $(\delta T/T)^3$, where $\delta T = T_{\rm top} - T_{\rm bot}$ is the temperature difference between the top and bottom viewports:
\begin{equation}
\begin{split}
    \Delta \nu_{{\rm BBR}, ji} &= -2 \frac{\alpha_{\rm Sr}\sigma_{\rm SB}}{h\epsilon_0 c}\frac{r^2}{4 Z_0^3}\delta z_{ji} T^4 \left[
    \left(1 + \frac{\delta T}{T}\right)^4 - 1
    \right]\\
    & \approx -2 \frac{\alpha_{\rm Sr}\sigma_{\rm SB}}{h\epsilon_0 c}\frac{r^2}{4 Z_0^3}\delta z_{ji} T^4 \left[
    4\left(\frac{\delta T}{T}\right) + 6\left(\frac{\delta T}{T}\right)^2 + 4 \left(\frac{\delta T}{T}\right)^3 +\mathcal{O}\left(\left(\frac{\delta T}{T}\right)^4\right)
    \right].
\label{BBR_expansion}
\end{split}
\end{equation}
Because the absolute temperature ($T=300$ K) is a factor of $150\times$ greater than the temperature difference we applied to the viewports ($\delta T\le2$ K),
the magnitude of the first-order term is roughly a factor of $100\times$ larger than the second-order term.
Therefore, we can approximate Eq.~\ref{BBR_expansion} as an expression with a linear scaling with both $\delta T$ and $\delta z_{ji}$:

\begin{equation}
    \Delta \nu_{{\rm BBR}, ji} \approx -2 \frac{\alpha_{\rm Sr}\sigma_{\rm SB}}{h\epsilon_0 c}\frac{r^2}{Z_0^3}T^3\times\delta z_{ji} \times \delta T . 
\label{BBR_final}
\end{equation}

Measuring the frequency differences of the 10 ensemble pairs simultaneously,
we find that as expected from Eq.~\ref{BBR_final}, the resulting frequency shifts scale linearly with the temperature difference, and the corresponding slopes scale roughly linearly with the height differences (Supplementary Figure~3).
Through a linear fit to the extracted slopes as a function of height differences,
we find the BBR sensitivity in our system to be $-4.2(1)\times10^{-18}/$cm per 1 K difference (Fig.~2c in the main text).

To monitor the temperature,
we use commercially available 10 k$\Omega$ negative temperature coefficient thermistors (Amphenol Thermometrics, MC65F103A),
rated for an interchangeability of 50 mK by the manufacturer.
Relative accuracy of the temperature sensors is calibrated in an ice water bath,
and we find the temperature differences between each sensor consistent within 25 mK.
The temperature sensors are mounted on the stainless steel Conflat flanges of the top and bottom viewports,
as well as the side flanges of the science chamber for monitoring the radial temperature inhomogeneity.
Under normal operating conditions,
we find the temperature difference between the top and bottom sensors lies within a range of $350\pm100$ mK and find the radial temperature inhomogeneity to be less than 75 mK. 

The MOT coils are mounted in the recessed window of the science chamber,
which could potentially introduce temporal thermal drifts during the sample preparation stage.
To suppress this,
we opt for compact coils with efficient water cooling through hollow wires,
keep the ${}^1S_0\leftrightarrow{}^1P_1$ first-stage MOT loading time ($<0.5$ s) short compared to the experimental cycle time ($12$ s),
and ensure that the duty-cycles remain consistent throughout the measurements.

While the strontium oven atom source (AOSense) is another potential source of BBR shift,
the atom ensembles have no direct line of sight to the BBR photons from the oven thanks to the use of a 2D-MOT for atomic beam deflection.
We operate the oven at the lowest possible temperature ($T=360~^\circ$C) under normal conditions,
and find no statistically significant change in the frequency differences across the ensemble array at the $1\times10^{-19}/$cm level when intentionally increasing the oven temperature to 460 $^\circ$C,
a lever arm of $1.8$ due to $T^{4}$ scaling bounds the systematic gradient from the oven to below $5.5\times10^{-20}/$cm.

Overall,
for measurements taken under normal operations,
we find the averaged temperature difference between top and bottom sensors to be $370(35)$ mK,
resulting in a BBR gradient of $-15.7(1.5)\times10^{-19}/$cm.

\clearpage

\captionsetup[figure]{labelfont={},name={Supplementary Figure}, labelsep=period}
 
\renewcommand{\thefigure}{3}
\begin{figure}[!ht]
    \centering
    \includegraphics[width=0.85\textwidth]{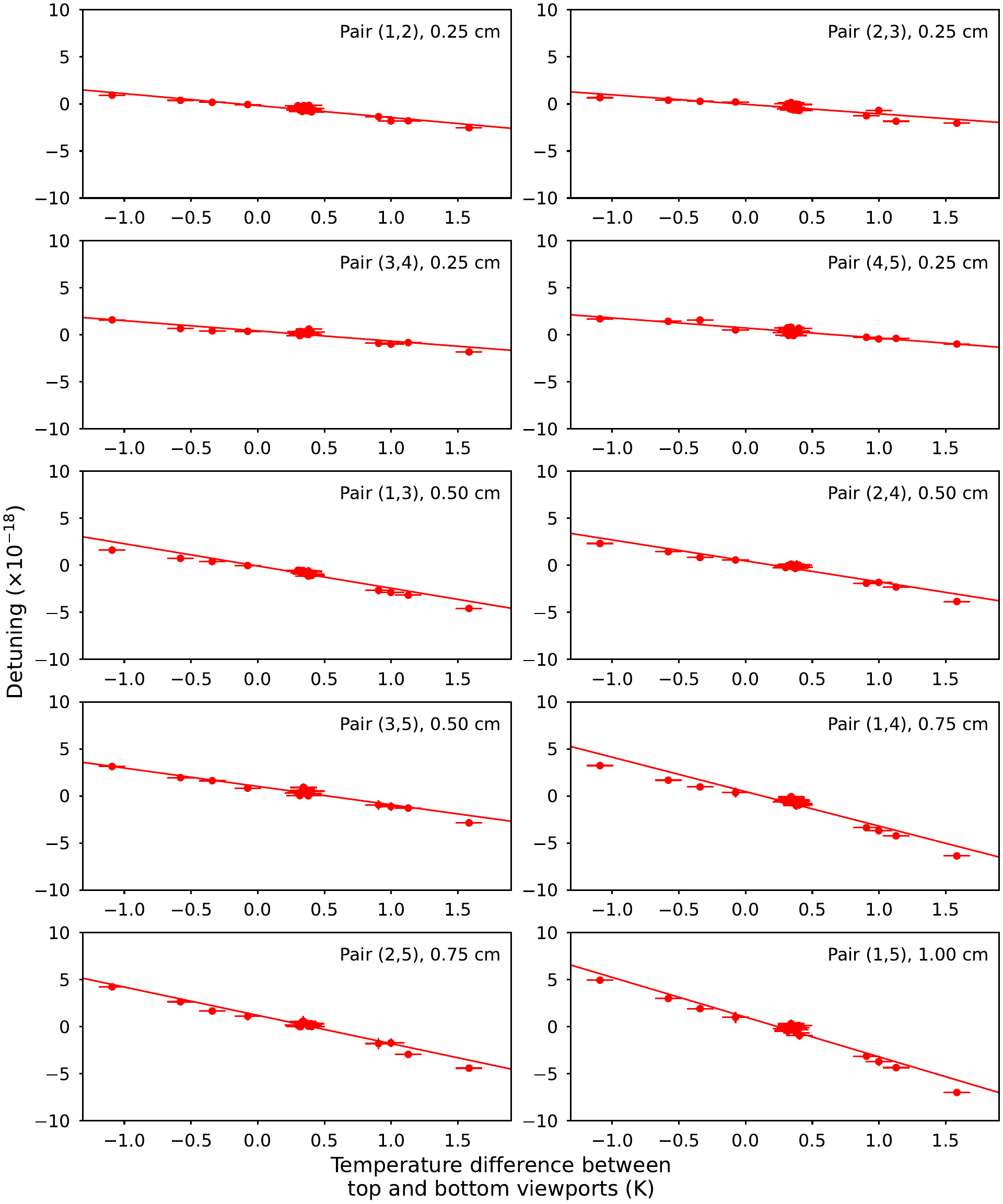}
    \caption{
    Black body radiation shift evaluation.
    Evaluations of BBR shift for all 10 ensemble pairs simultaneously.
    Solid lines are linear fits to the scatter points.
    The y-axis scales are kept the same for all the plots.
    The dense data points around 0.35 K correspond to temperature differences recorded under normal operating conditions.
    The inset text list the ensemble pair indices and corresponding height difference.
    }
\end{figure}

\subsection{Lattice light shift}

The lattice light shift for a single atomic ensemble is given by \cite{ushijima_operational_2018s}:
\begin{equation}
\begin{split}
    \nu_{LS}(u, \delta_L, n_z) \approx &\bigg( \frac{\partial \tilde\alpha^{E1}}{\partial \nu} \delta_L - \tilde\alpha^{qm} \bigg)\bigg( n_z + \frac{1}{2}\bigg) u^{1/2}\\
    & -\bigg[\frac{\partial\tilde \alpha ^{E1}}{\partial \nu}\delta_L + \frac{3}{2}\tilde\beta\bigg(n_z^2 + n_z+\frac{1}{2} \bigg) \bigg]u \\
    & +2\tilde\beta \bigg(n_z +\frac{1}{2} \bigg)u^{3/2} -\tilde\beta u^2,
\end{split}
\label{LS_abs}
\end{equation}
in which $\tilde\alpha^{E1}$, $\tilde\alpha^{qm}$, and $\tilde \beta$ are the differential $E$1, $E$2-$M$1 polarizabilities, 
and hyperpolarizability on the clock transition, respectively.
$u$ is the lattice trap depth in units of $E_{\textrm{rec}}$,
$n_z$ is the axial vibrational quanta,
and $\delta_L$ is lattice detuning from the effective magic wavelength, where the scalar and tensor shifts cancel.
The vector shift is rejected by averaging the transitions with opposite nuclear spin states.
Through clock sideband thermometry both axially and radially
we find $n_z <0.01$ and radial temperature $T_{r}<200$ nK at $u_{op} = 15~E_{\rm rec}$,
where thermal averaging of the effective trap depth can be neglected.
By operating at lattice depths below $35~E_{\rm{rec}}$,
the hyperpolarizability terms are also negligible.

To model the differential lattice light shift in our system,
we introduce a dimensionless parameter $\delta u$,
which characterizes the relative lattice trap depth difference between ensemble pair $(i, j)$, where $j>i$
\begin{equation}
\begin{split}
    u_i &=u,\\
    u_j &=u(1+\delta u).
\end{split}
\end{equation}

For $\delta u \ll 1$,
the differential light shift ($\nu_{LS, j} -\nu_{LS, i}$) can be approximated as
\begin{equation}
    \Delta\nu_{LS}(\delta_L, \delta u, u) =\delta u\bigg[\bigg(\frac{\partial\tilde\alpha^{E1}}{\partial \nu}\delta_L - \tilde\alpha^{qm}\bigg) \frac{1}{4}\sqrt{u} - \frac{\partial\tilde\alpha^{E1}}{\partial\nu}\delta_L u\bigg],
\label{dLS1}
\end{equation}
which scales linearly with $\delta u$.

By modulating the lattice detuning between $\delta_L +100$ MHz and $\delta_L - 100$ MHz at the operational depth $u_{op}=15~E_{\rm rec}$, we have
\begin{equation}
\begin{split}
    &\Delta\nu_{LS}(u_{op}, \delta_L + 100~{\rm MHz}) -\Delta\nu_{LS}(u_{op}, \delta_L - 100~{\rm MHz})\\ 
    =&~ \delta u \frac{\partial\tilde\alpha^{E1}}{\partial \nu} \bigg(\frac{1}{4}\sqrt{u_{op}} - u_{op} \bigg)\times 200~{\rm MHz},
\end{split} 
\end{equation}
allowing extraction of $\delta u$,
which are found to be below 5 \% for all the ensemble pairs,
and are symmetric around 0 (Supplementary Figure~4a),
reflecting the Gaussian nature of the lattice beam profile and that the array is centered about the focus.
Through mapping out the frequency shifts for 10 ensemble pairs when modulating the lattice intensity,
we find the differential light shifts scale with the spatial separations between ensemble pairs but do not scale with $\delta u$,
resulting in a residual spatial light shift gradient of $-8.0(1.1)\times10^{-20}/E_{\rm rec}/$cm~(Supplementary Figure~4b).

Although the origin of the spatial light shift gradient is not definitively known,
we also observe a differential lattice vector Stark gradient of $-2.5(2)\times10^{-18}/E_{\rm rec}/$cm by measuring the splitting between transitions with opposite spin states~(Supplementary Figure~4c),
which also does not scale with $\delta u$ and instead scales with trap depth and spatial separation.
For a single ensemble,
the vector Stark shift is proportional to $\xi_{\rm lat}( \bf{e_k}\cdot \bf{e_B})$,
where $\xi_{\rm lat}$ is the ellipticity of the lattice light,
$\bf{e_k}$ is the lattice wave vector along $\bf{z}$ direction,
and $\bf{e_B}$ is the magnetic field vector primarily along $\bf{x}$.
We find the vector Stark gradient arises from the spatially varying coefficient,
$\bf{e_k}\cdot \bf{e_B}$,
verified through the change of vector Stark gradient by applying an additional magnetic field gradient $\partial B/\partial z$ of up to $\pm 10$ mG$/$cm,
which effectively changes $(\bf{e_k}\cdot \bf{ e_{B,j}} - \bf{e_k}\cdot\bf{e_{B,i}})$.
Based on this observation,
we hypothesize that the residual spatial light shift gradient is due to a differential tensor Stark shift,
in which the spatially varying magnetic field vectors are coupled to the (nearly) linear lattice polarization ($\bf{ e_{\rm lat} }$, along $\bf{ x}$) through $|\bf{ e_{\rm lat}}\cdot \bf{e_B} |^2$.
The differential vector Stark shift results in a gradient of $-37(3)\times10^{-18}/$cm at $u_{op}=15~E_{\rm rec}$ in the splitting between transitions with opposite spin states,
equivalent to a fictitious magnetic field gradient of $\sim600~\mu$G$/$cm and is accounted for during the second-order Zeeman shift corrections.

To independently evaluate the light shift gradient,
we need to account for the $\delta u$ dependent shifts,
which require the knowledge of $\delta u$ and $\delta_L$.
The latter is done by first subtracting the residual spatial gradient from the measured differential light shifts. We then find that the remaining shifts are correlated with the extracted $\delta u$ (Fig.~2d in the main text) as expected from Eq.~\ref{dLS1}.
From this we find our operational lattice detuning $\delta_L$ to be $16.9(1.5)$ MHz.
With the extracted $\delta_L$ and $\delta u$ calibrated from each measurement day,
we are able to monitor and account for the residual light shift gradient.
A representative plot is shown in Fig.~2e in the main text.

Our operational lattice frequency is measured to be $368,554,780(30)$ MHz,
limited by the accuracy of the wave-meter~(HighFinesse, WS7).
However,
the lattice frequency is stabilized to a ultra-low-expansion cavity with typical drifts of less than $-20$ kHz per day,
inferred from the narrow-linewidth ${}^1S_0\leftrightarrow{}^3P_1$ MOT,
providing sufficient long-term stability throughout the measurements.
The lattice intensity is actively stabilized and controlled via feedback to the acoustic-optical modulator before the beam delivering optical fiber (NKT Photonics, LMA-PM-15).
Both the incoming and retro-reflected lattice alignments are monitored on several cameras and photo-diodes,
ensuring the daily-calibrated $\delta u$ remained symmetric around 0 throughout the data taking campaign.
Overall, the lattice light shift gradient in our system is evaluated to be $-11.8(1.2)\times10^{-19}/$cm.

\captionsetup[figure]{labelfont={},name={Supplementary Figure}, labelsep=period}
 
\renewcommand{\thefigure}{4}
\begin{figure}[!ht]
    \centering
    \includegraphics[width=1.0\textwidth]{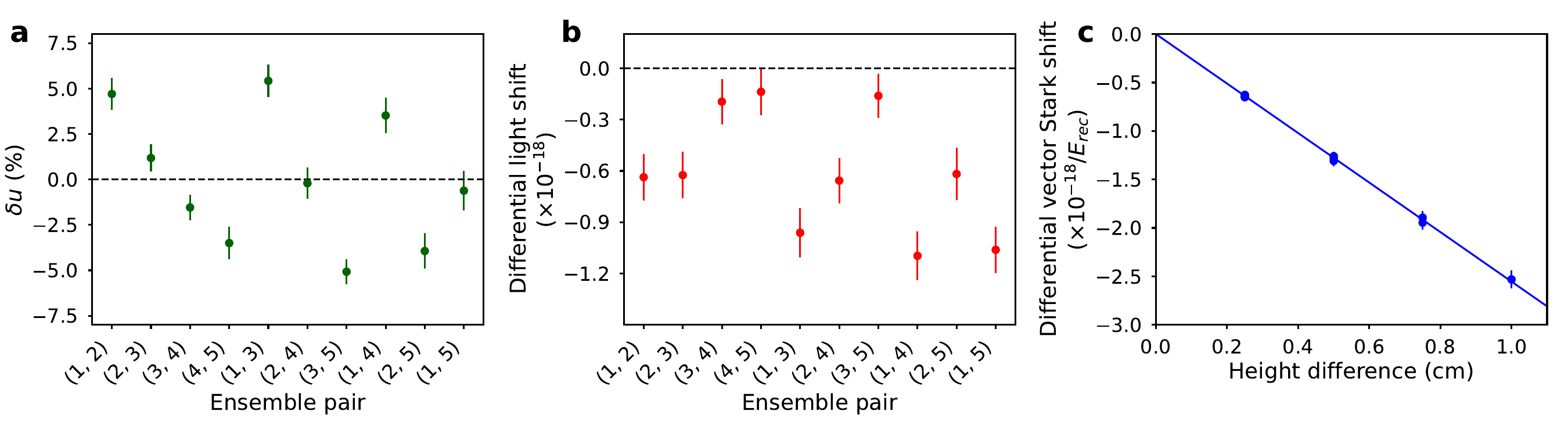}
    \caption{
    Lattice light shift evaluation.
    (a) Extraction of relative trap depth difference $\delta u$ for 10 ensemble pairs.
    The extracted $\delta u$ are symmetric around 0,
    reflecting the Gaussian nature of the lattice beam profile and that the ensembles are centered about the beam focus.
    (b) Evaluation of differential lattice light shifts at $u_{op}=15~E_{\rm rec}$ for 10 ensemble pairs.
    The shifts scale with spatial separation,
    but do not scale with $\delta u$,
    resulting in a residual spatial lattice light shift gradient of $-8.0(1.1)\times10^{-20}/E_{\rm rec}/$cm.
    (c) Evaluation of the differential vector Stark shift,
    which also doesn't scale with $\delta u$.
    A linear fit (blue solid line) yields a gradient of $-2.5(2)\times10^{-18}/E_{\rm rec}/$cm.
    }
\end{figure}

\subsection{Probe Stark shift}

The probe AC Stark shift arises from the clock light itself,
and is suppressed when probing with a shared clock light,
with uncertainties primarily arising from inhomogeneity and misalignment of the clock beam with respect to the lattice.
The clock light beam waist is $\sim1$ mm,
a factor of 10 greater than that of the lattice,
ensuring homogeneity across the
atomic ensembles both axially and radially.
The clock light is carefully aligned to the lattice light,
and is monitored using several cameras by picking-off the beam before and after the science chamber over a distance of $\sim2.5$ m.
The alignment is further verified by ensuring that the time periods of the clock transition Rabi oscillations of all 5 ensembles agree within $1\%$ fitting uncertainty before and after each measurement run.

\captionsetup[figure]{labelfont={},name={Supplementary Figure}, labelsep=period}
 
\renewcommand{\thefigure}{5}
\begin{figure}[!ht]
    \centering
    \includegraphics[width=0.75\textwidth]{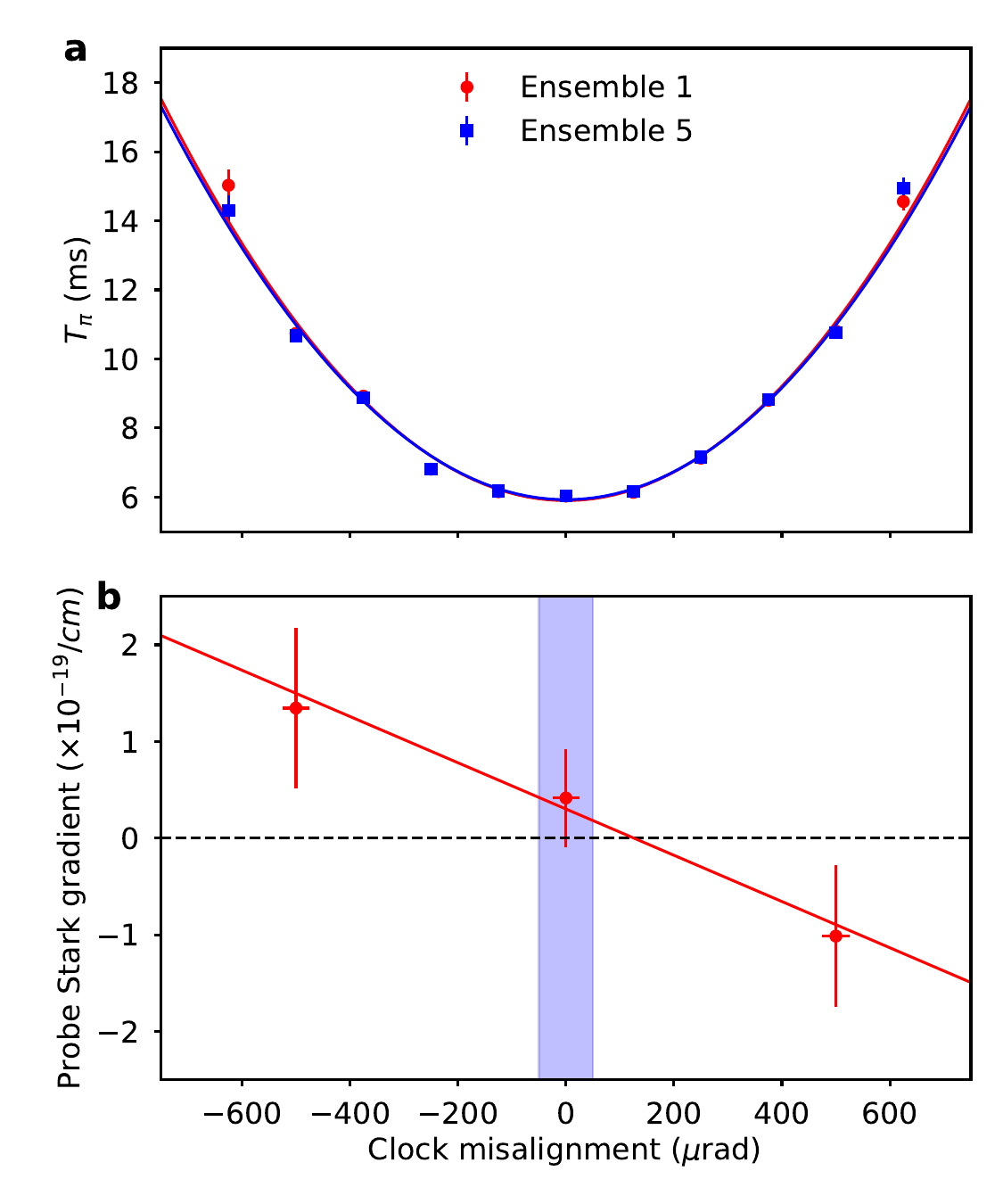}
    \caption{
    Probe Stark shift evaluation.
    (a) Measured $\pi$ pulse durations $T_{\pi}$ for ensemble 1 (red) and 5 (blue), the pair with largest spatial separation (1 cm),
    as a function of the angle of clock beam misalignment.
    The solid lines are quadratic fit to the data.
    (b) Evaluated probe Stark shift at the operational intensity $I_{op}$ at $\pm 500~\mu$rad clock beam misalignment (estimated using camera).
    A linear fit to the data bounds the probe Stark shift gradient to be below $0.5\times10^{-19}/$cm (the blue area represents a $\pm 50~\mu$rad range).
    }
\end{figure}

Evaluation of the probe Stark effect is performed by interleaving between the operational clock intensity of $I_{op}$,
corresponding to a $\pi$ pulse duration of 6 ms,
and the case of $4 I_{op}$.
This interleaved measurement is repeated by further misaligning the clock beam by up to $\pm500~\mu$rad (estimated via camera images),
at which the $\pi$ pulse durations increase by about a factor of 2 (Supplementary Figure~5).
Through linear fitting,
we find a frequency gradient of $2.5(1.0)\times10^{-19}/$cm across the span of $\pm500~\mu$rad misalignments.
Under normal operations,
the misalignment is monitored and bounded within a $\pm 50~\mu$rad range which accounts for possible drifts during the experiments.
This bounds the uncertainty from probe Stark shift to below $0.5\times10^{-19}/$cm.

\subsection{DC Stark shift}

Frequency shifts arising from electric fields can perturb the clock transition through DC Stark effect of the form $kE^2$,
where $E={\rm|\bf E|}$ is the static electric field and $k\equiv-(\alpha_e - \alpha_g)/(2h)$ is the coefficient specific to the clock transition,
with $\alpha_{g,e}$ being the static polarizabilities of the ground and excited clock states.
To evaluate the DC Stark effect,
a pair of quadrant electrodes are mounted along the lattice axis outside the top and bottom science chamber viewports,
with a total separation of roughly 30 cm.
We then probe the clock transition with opposite voltages applied to the electrode pair and interleave between ($+V,-V$) and ($-V,+V$) configurations. We observe differential shifts below the $1\times10^{-19}/$cm level when applying voltages of up to $\pm 100$ V to the electrodes, indicating that the background electric field gradient is small.

To quantify our uncertainty in the differential shifts due to the background electric field gradient,
we follow the approach laid out in Ref.~\cite{beloyFaradayShielded_2018s}. In the presence of a background field $E_{bg}$, which would most likely arise due to charge accumulation on the nearby top and bottom viewports,
the DC Stark shift for a single atom ensemble is given by $k(E_{bg} + cV)^2$,
where $c$ is the atomic coefficient.
We rewrite the above form as~\cite{beloyFaradayShielded_2018s}
\begin{equation}
    \nu_{dc}(V) = \nu_{bg} + a V + b V^2,
\end{equation}
where $\nu_{bg}$ is the background stray field shift,
and coefficients $a$ and $b$ are experimentally accessible parameters by modulating $V$.
We note that $aV$ represents the coupling between the background field and the applied field.
When comparing ensemble $i$ and ensemble $j$ ($j>i$),
the differential DC Stark shift ($\Delta\nu_{dc} = \nu_j -\nu_i $) becomes
\begin{equation}
    \Delta\nu_{dc}(V) = \Delta\nu_{bg} + A V + B V^2,
    \label{DCStark}
\end{equation}
where $A = a_j - a_i$,
$B = b_j - b_i$,
and $\Delta\nu_{bg} = \nu_{bg,j}-\nu_{bg, i}$ is the differential background shift due to charges on the viewports.

Due to the finite spatial extent ($s$) of the atom ensembles and inhomogeneity of the applied field,
there is no $V$ such that the applied field identically cancels $\Delta\nu_{bg}$.
We consider the extremum value of $\Delta\nu_{dc}(V) $,
denoted as $\Delta\nu^*$.
We then have the difference between the differential background shift and the extremum shift, $\Delta\nu_{\rm diff}\equiv\Delta\nu_{bg}-\Delta\nu^*$ (refer to Fig.~1 in Ref.~\cite{beloyFaradayShielded_2018s} for details).
From Eq.~\ref{DCStark},
we find $\Delta\nu_{\rm diff} = A^2/(4B)$.
We note that $\Delta\nu^*$ plays the role of a frequency correction for the field gradients,
and can be written as $\Delta\nu^* = \eta \Delta\nu_{\rm diff}$, due to the fact that $\Delta\nu^*$ and $\Delta\nu_{\rm diff}$ scale similarly with the stray field.
The differential background shift is then given by
\begin{equation}
    \Delta\nu_{bg} = (1+\eta)\Delta\nu_{\rm diff} = (1+\eta)A^2/(4B),
    \label{diffDCStark}
\end{equation}
in which $\eta$ is given by $\eta = \zeta^2s^2/R^2$ to leading order of $s$,
where $\zeta\equiv(q_1+q_2)/(q_1-q_2)$ quantifies the charge symmetry between the viewports,
and $R$ is the effect length.

By applying voltages of up to $\pm 100$ V to the electrodes,
we find $A^2/(4B) = 0.6(3)\times10^{-20}$ for the ensemble pair of 1 cm separation.
In our system,
we obtain $\eta \approx 0.02$ for $s=~500~\mu$m,
$R=40$ mm,
and assuming $25\%$ more charge on one viewport than the other.
This bounds the background DC Stark shift gradient to below $0.1\times10^{-19}/$cm.

\subsection{Ellipse fitting}

In the presence of QPN and numerical constraints which ensure an ellipse-specific solution,
the least squares ellipse fitting approach is biased at phases close to 0 or $\pi$ where the ellipse collapses into a straight line~\cite{ellipse_fitting_s}.
The differential phases across the ensemble array are dominated by the differential Zeeman shifts,
such that through precise control of the magnetic field and gradient,
we can operate in the regime where all the differential phases lie within the $(\pi/6, 5\pi/6)$ range.
For typical experiments with a 10 s Ramsey dark time and 2000 atoms per ensemble,
the ellipse fitting bias error is bound to below $2\times10^{-19}$~(Supplementary Figure~6).
During initial data processing,
the bias error is corrected using Monte-Carlo simulations with the experiment parameters such as atom number and contrast as inputs.
While the bias error does not scale with spatial separations,
we estimate an upper bound limit of $0.5\times10^{-19}$ due to uncertainty in the input parameters used for Monte-Carlo simulation during bias error corrections.

\captionsetup[figure]{labelfont={},name={Supplementary Figure}, labelsep=period}
 
\renewcommand{\thefigure}{6}
\begin{figure}[!ht]
    \centering
    \includegraphics[width=0.85\textwidth]{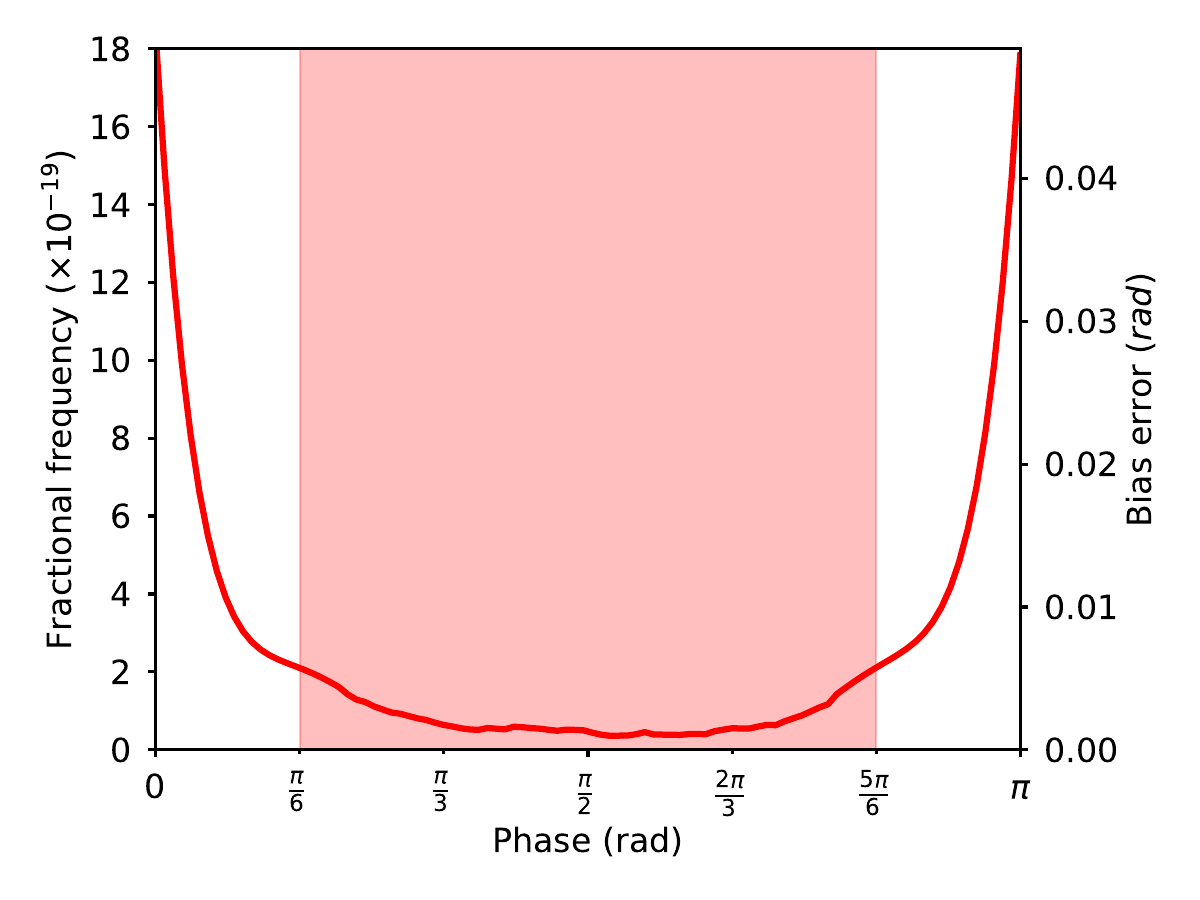}
    \caption{
    Ellipse fitting bias error evaluation.
    Monte-Carlo simulations of bias error from ellipse fitting with input parameters of 2000 atoms per ensemble, 85\% contrast and 10 s Ramsey dark time (red solid line).
    The red area represents our normal operating range of $(\pi/6, 5\pi/6)$,
    where fractional bias error is bound below $2\times10^{-19}$ and is corrected for each ellipse fitting during data analysis with uncertainty below $0.5\times10^{-19}$.
    }
\end{figure}

\subsection{Wannier-Stark ladder, imaging, and the expected gravitational redshift}

The local gravitational acceleration in our laboratory was measured
by the group of Prof.~Tikoff from the Department of Geoscience at the University of Wisconsin-Madison using a LaCoste-Romberg gravimeter,
and was then cross-validated with known values from several other survey points in Wisconsin.
The local gravitational acceleration in our laboratory was measured to be $g = -9.803$ m$/$s$^2$, rounded to the 4$^{\rm th}$ digit.

The lattice tilt ($\theta_{\rm tilt}$) with respect to gravity is independently measured using the splitting of the $\pm1^{\rm st}$ order Wannier-Stark sidebands~\cite{lemonde_WS_2005s} to the clock transition at $5~E_{\rm rec}$ lattice depth (Supplementary Figure~7a),
which is given by
\begin{equation}
    \Delta\nu_{\rm WS, \pm1} = mg\lambda_L{\rm cos}\theta_{\rm tilt},
\end{equation}
where $m$ is the mass of ${}^{87}$Sr,
and $\lambda_L=813.4$ nm is the lattice wavelength.
A representative plot of the measured Wannier-Stark sidebands is shown in Supplementary Figure~7b,
and we find the weighted average splitting between $\pm1^{\rm st}$ order sidebands to be 1729(2) Hz,
corresponding to a tilt with respect to gravity of $5.0(1)^\circ$.

The overall spatial extent of the ensemble array is $1.00(1)$ cm,
calculated based on the frequency chirp profile of the moving optical lattice loading sequence,
which is precisely controlled using a direct digital synthesizer (DDS, Moglabs XRF).
This yields an effective height difference of $\Delta h = 0.99(1)$ cm,
consistent with the extraction of the height difference from the camera images.
The effective camera pixel size in our imaging system is calibrated to be $34(1)~\mu$m per pixel using the standard time-of-flight imaging.

Fluctuations of background magnetic fields modifying the narrow-line MOT operation could result in position drift in the atom ensemble.
This is rejected by identifying the center-of-mass of each ensemble with a 2D Gaussian fit,
followed by a selection of region-of-interest (ROI) with a 15 pixel by 15 pixel region centered on the fit.
Each ensemble has a finite vertical spatial extent of roughly $500(10)~\mu$m,
corresponding to an expected gradient of $-5.5(1)\times10^{-20}$ across the ensemble due to the gravitational redshift,
which is averaged out by spatial averaging over the ROI in order to extract the redshifts between the center-of-mass positions of each ensemble.
The center-of-mass position uncertainty of each ensemble is limited by the DDS timing accuracy ($100$ ns), corresponding a position error of $<0.1~\mu$m.

The expected gravitational redshift from theory of general relativity at a 1 cm height difference is then given by
\begin{equation}
    \frac{g\Delta h}{c^2} = -10.9 \times10^{-19},
\end{equation}
where $c$ is the speed of light.
This results in an expected redshift gradient of $-10.9\times10^{-19}/$cm,
with the uncertainty bounded $<0.1\times10^{-19}/$cm due to uncertainties in $g$ and $\Delta h$.

\newpage

\captionsetup[figure]{labelfont={},name={Supplementary Figure}, labelsep=period}
 
\renewcommand{\thefigure}{7}
\begin{figure}[!ht]
    \centering
    \includegraphics[width=0.95\textwidth]{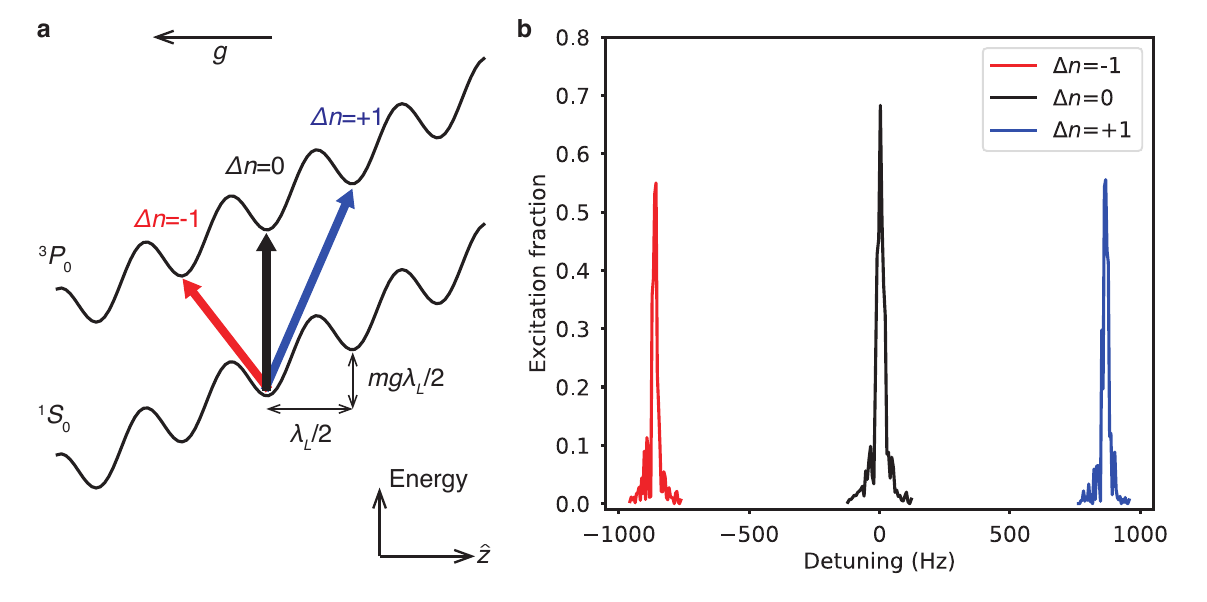}
    \caption{
    Wannier-Stark ladder sidebands.
    (a) Illustration of the Wannier-Stark ladder in a 1D vertical optical lattice along the $\bf{z}$ direction.
    The Wannier state at lattice site $n$ has energy $nmg\lambda_L/2$.
    The black arrow represents the carrier clock transition where $n$ remains unchanged ($\Delta n = 0$).
    The blue (red) arrow represents the first-order sideband transition with $\Delta n = +1 (-1)$.
    (b) Representative plot of measured Wannier-Stark sidebands at a shallow lattice depth ($5~E_{\rm rec}$) recorded with Fourier-limited Rabi spectroscopy with a $\pi$ pulse duration of 40 ms.
    The black line corresponds to the carrier ($\Delta n=0$) transition,
    and the blue (red) line corresponds to off-site transitions, $\Delta n = +1 (-1)$.
    The splitting between $\Delta n=\pm1$ transitions is used to extract the lattice tilt.
    }
\end{figure}

\newpage

\captionsetup[figure]{labelfont={},name={Supplementary Figure}, labelsep=period}
 
\renewcommand{\thefigure}{8}
\begin{figure}[!ht]
    \centering
    \includegraphics[width=0.95\textwidth]{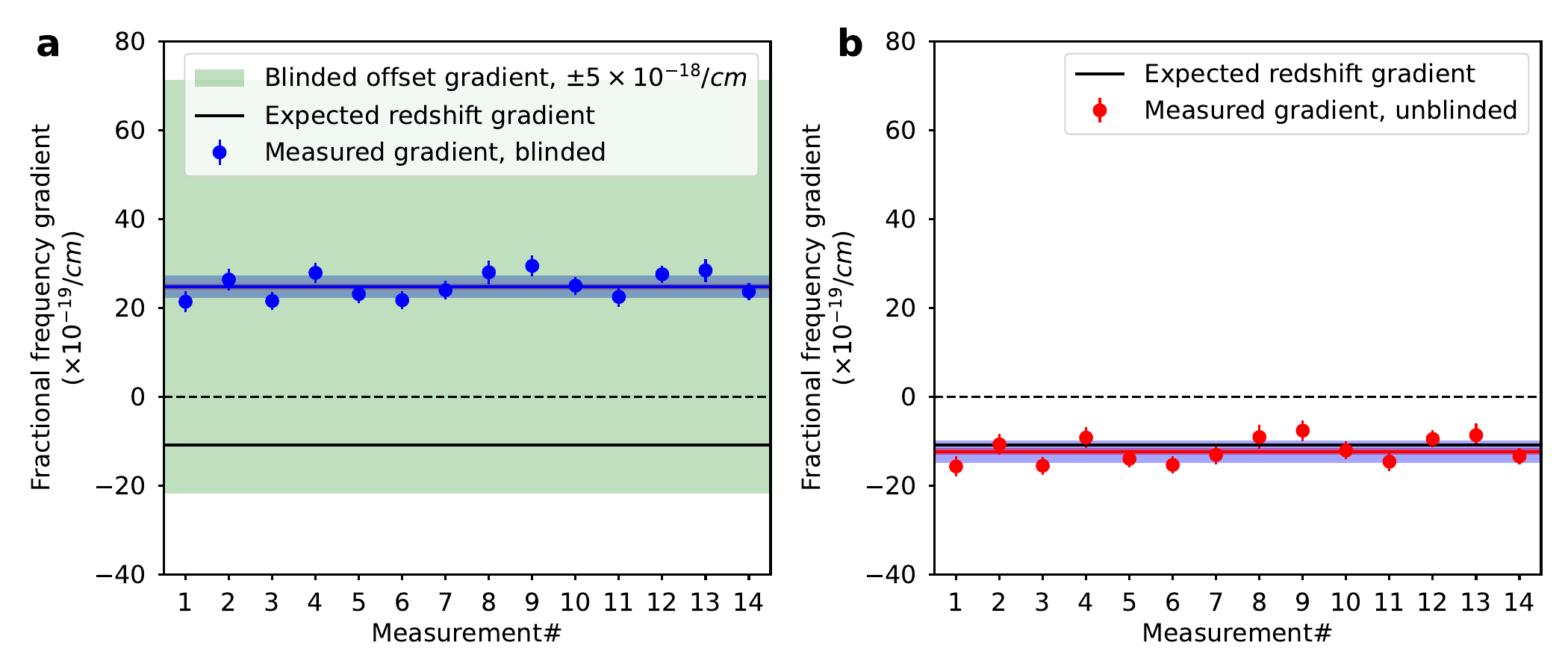}
    \caption{
    Measurements before and after removal of the blinded offset gradient.
    (a) Measured fractional frequency gradients (blue scatter points) with blinded offset. 
    The blue area represents $\pm1\sigma$ total uncertainty.
    The black solid line is the expected gravitational redshift gradient of $-10.9\times10^{-19}/$cm. 
    The green area represents the range of possible outcomes upon removal of the blinded offset gradient,
    corresponding to a $\pm5\times10^{-18}/$cm range. All corrections and uncertainties were finalized at this stage, without knowledge of the blinded offset. 
    (b) Measured fractional frequency gradients following removal of the blinded offset. Red scatter points are the same data set as in (a) with the blinded offset gradient ($+3.7\times10^{-18}/$cm) removed.
    This represents the same data shown in Fig.~3a in the main text, but here the scale of the y-axis is kept the same as in panel (a) for the sake of comparison.
    }
\end{figure}

\newpage

\captionsetup[figure]{labelfont={},name={Supplementary Table}, labelsep=period}
 
\renewcommand{\thefigure}{1}
\begin{figure}[!ht]
    \centering
    \includegraphics[width=0.95\textwidth]{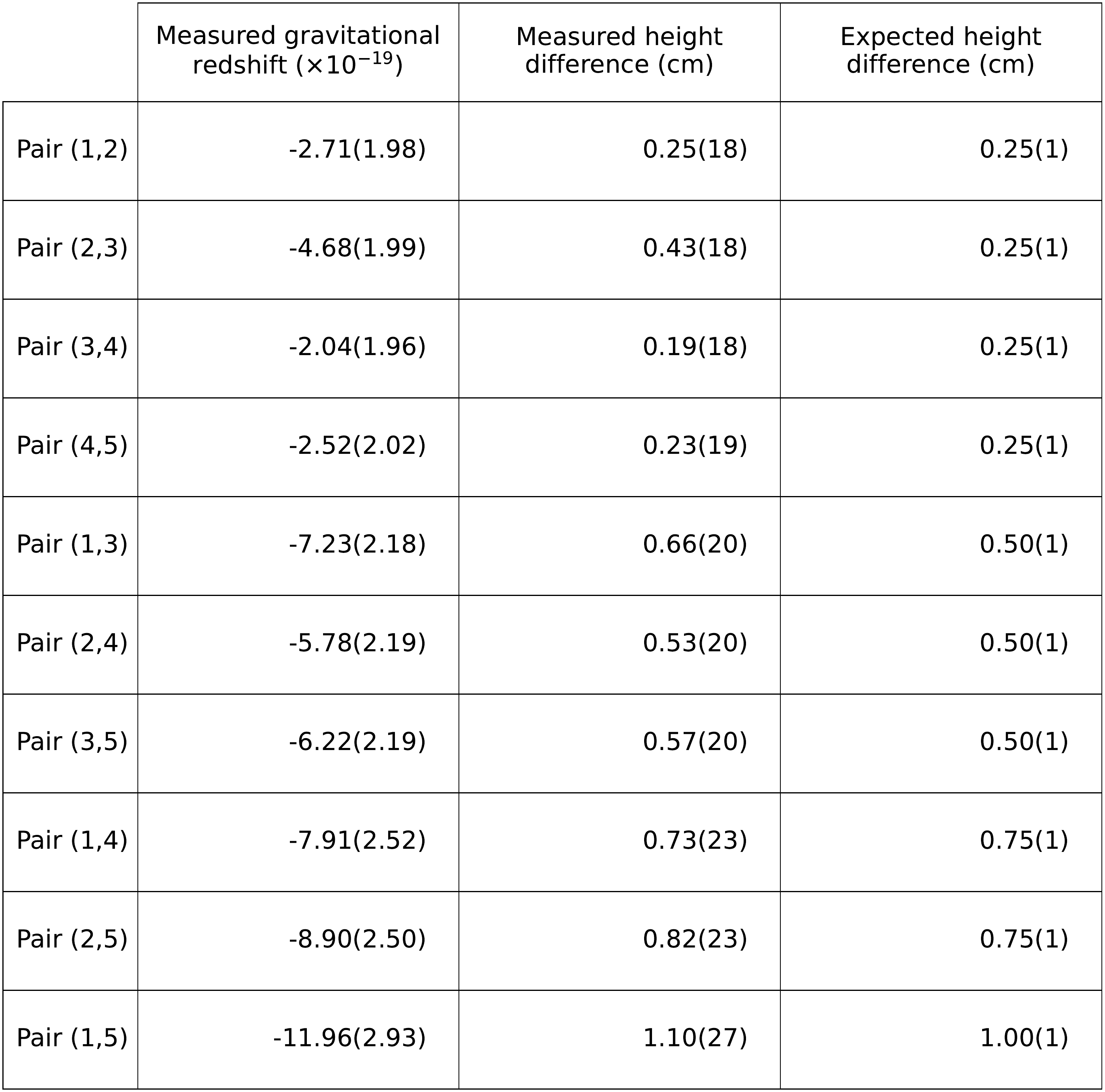}
    \caption{
    Clock height difference measurement.
    The measured gravitational redshift of each clock comparison is used to extract the height difference given the independently measured local gravitational acceleration, $g=-9.803$ m$/$s$^2$.
    }
\end{figure}

\end{document}